\documentclass[10pt,final,twoside,journal]{IEEEtran}

\usepackage{amsmath}
\usepackage{amssymb}
\usepackage{amsfonts}
\usepackage{amscd}
\usepackage{amsthm}
\usepackage{mathrsfs}
\usepackage[noadjust]{cite}
\usepackage{threeparttable}
\usepackage{textcomp}
\usepackage{multirow}

\newtheorem{theorem}{Theorem}

\newtheorem{proposition}{Proposition}

\theoremstyle{remark}
\newtheorem{example}{Example}

\makeatletter
\renewcommand*\env@matrix[1][c]{\hskip -\arraycolsep
  \let\@ifnextchar\new@ifnextchar
  \array{*\c@MaxMatrixCols #1}}
\makeatother
\usepackage{graphicx}

\usepackage{etoolbox}
\AtEndEnvironment{example}{\null\hfill\IEEEQED}%

\newcommand{\Hsub}{{\bf H}_{\rm sub}}
\newcommand{\musub}{\mu_{\rm sub}}

\usepackage{paralist}

\title{Line of Sight $2 \times n_r$ MIMO \\ with Random Antenna Orientations}
\author{
\IEEEauthorblockN{Lakshmi~Natarajan, Yi~Hong,~\IEEEmembership{Senior~Member,~IEEE}, and Emanuele~Viterbo,~\IEEEmembership{Fellow,~IEEE}}
\thanks{Copyright \copyright 2015 IEEE. Personal use of this material
is permitted. However, permission to use this material for any other purposes must be obtained from
the IEEE by sending a request to pubs-permissions@ieee.org.}%
\thanks{This work was supported by Australian Research Council Discovery Project under Grant ARC~DP160100528.}%
\thanks{L.~Natarajan is with the Department of Electrical Engineering, Indian Institute of Technology Hyderabad, Sangareddy 502285, India (email: lakshminatarajan@iith.ac.in).}%
\thanks{Y.~Hong and E.~Viterbo are with the Department of Electrical and Computer System Engineering, Monash University, VIC 3800, Australia (e-mail: \{yi.hong, emanuele.viterbo\}@monash.edu).}%
}

\begin{document}

%% \IEEEpubid{0000--0000/00\$00.00~\copyright~2015 IEEE}

\maketitle

%%%%%%%%%%%%%%%%%

\begin{abstract}
\boldmath
Line-of-sight (LoS) multiple-input multiple-output (MIMO) gives full spatial-multiplexing gain when the antenna array geometry and orientation are designed based on the inter-terminal distance. These known design methodologies, that hold for antenna arrays with fixed orientation, do not provide full MIMO gains for arbitrary array orientations. In this paper, we study LoS MIMO channels with random array orientations when the number of transmit antennas used for signalling is $2$.
We study the impact of common array geometries on error probability, and identify the code design parameter that describes the high signal-to-noise ratio (${\sf SNR}$) error performance of an arbitrary coding scheme. For planar receive arrays, the error rate is shown to decay only as fast as that of a rank~$1$ channel, and no better than ${\sf SNR}^{-3}$ for a class of coding schemes that includes spatial multiplexing. We then show that for the tetrahedral receive array, which uses the smallest number of antennas among non-planar arrays, the error rate decays faster than that of rank~$1$ channels and is exponential in ${\sf SNR}$ for every coding scheme. Finally, we design a LoS MIMO system that guarantees a good error performance for all transmit/receive array orientations and over a range of inter-terminal distances.
\end{abstract}


\section{Introduction} \label{sec:1}

\IEEEPARstart{T}{he large} swathes of raw spectrum available in the millimeter-wave frequency range are expected to provide an attractive solution to the high data-rate demands of the future 5G cellular networks~\cite{RSM_Access_13}.
The small carrier wavelength of millimeter-wave frequencies allow for reduced spacing between the antenna elements when multiple antennas are used at the transmitter and receiver. This implies that multiple-input multiple-output (MIMO) spatial multiplexing gains can be obtained even in the presence of a strong line-of-sight~(LoS) component when operating in such high frequencies~\cite{TMR_JWCOM_11}.

In LoS environments, the MIMO channel matrix ${\bf H}$ is a deterministic function of the positions of the transmitter and receiver and the geometry of the antenna arrays used at either terminals. If the positions of the communicating terminals are fixed and known apriori, the geometry of the antenna arrays can be designed to optimize the performance of the communication system. 
The LoS MIMO channel quality, in terms of capacity, multiplexing gain, coverage and channel eigenvalues, 
have been studied in~\cite{BOO_WCNC_05,BOO_JWCOM_07,BOO_Eur_07,SaN_JVT_07,TMR_JWCOM_11,CeO_WCSP_13,WLYSV_JWCOM_14,WLYSV_ICC_14} as a function of the inter-terminal distance and the inter-antenna spacing of transmit and receive arrays, when the antennas are to be arranged in a rectangular, circular or a linear array. 
However, these design techniques assume that the position and the orientation of the antenna arrays are fixed, and the resulting criteria may be difficult to be satisfied if either of the communicating terminals is mobile or if the positions of the wireless terminals are not known a priori. Systems designed according to these known criteria degrade gracefully with variations in the geometric parameters, and may be adequate in certain scenarios where the changes in the orientation are limited, such as in a sectored communication cell where the variation of the base station orientation with respect to the direction of propagation is limited. 
However, these designs, which utilize two-dimensional antenna arrays, do not provide MIMO spatial multiplexing gains for arbitrary array orientations. 

%% \IEEEpubidadjcol

In~\cite{HKL_GLOBECOM_10}, the mutual information rates of a predominantly LoS channel with arbitrary antenna array orientations were studied using simulations and direct measurements in an indoor environment. The results show that the three-dimensional antenna arrays obtained by placing the antennas on the faces of a tetrahedron or a octahedron provide mutual information rates that are largely invariant to the rotation of antenna arrays in indoor LoS conditions. 
Previous studies of three-dimensional antenna arrays for wireless communications have mainly studied the capacity of the resulting MIMO system in a rich scattering environment. In~\cite{GeA_TWC} a compact MIMO antenna was proposed which consists of $12$ dipole antennas placed along the edges of a cube. A $24$-port and a $36$-port antenna were designed in~\cite{CYM_AP} by placing antennas along the edges and faces of a cube. In~\cite{LSK_ICATC} and~\cite{PaC_AWPL}, $6$-port and $16$-port antennas were designed on a cube, respectively, and the performance of the MIMO system in terms of capacity and channel eigenvalues in a richly scattering environment were studied. 
The objective of~\cite{GeA_TWC,CYM_AP,LSK_ICATC,PaC_AWPL} has been to design a compact array by densely packing the antenna elements while exploiting the degrees of freedom available in an environment that provides abundant multipath components.

%% Contributions %% 

To the best of our knowledge, there has been no prior theoretical study of LoS MIMO channels where the transmit or receive antenna array orientations are arbitrary, as may be experienced in wireless mobile communications. Further, all previous work have focussed on optimizing the mutual information rates of the MIMO channel. In order to achieve the information theoretic limits, we need code design criteria based on an error performance analysis of the communication channel.
In this paper, we consider LoS MIMO channels where the number of transmit antennas used for signalling is $2$ and both the transmit and receive arrays have random orientations. We study the impact of the geometry of the antenna arrays on the system error performance and design a LoS MIMO system that guarantees a minimum channel quality and good error performance for arbitrary transmit and receive orientations over a range of inter-terminal distances.

We model the $2$-transmit antenna $n_r$-receive antenna LoS MIMO channel ${\bf H}$ using the upper triangular matrix ${\bf R}$ obtained from its QR-decomposition~(Section~\ref{sec:2}). This allows us to derive bounds on pairwise error probability and identify the code parameter that determines the high signal-to-noise ratio (${\sf SNR}$) error performance of arbitrary coding schemes in LoS MIMO channels.

We show that for any planar, i.e., $2$-dimensional, arrangement of receive antennas (such as linear, circular and rectangular arrays), the rate of decay of error probability is similar to that of a rank~$1$ LoS MIMO channel whenever the receiver undergoes random rotations. Further, for some coding schemes, including \emph{spatial multiplexing}~\cite{Fos_Bell_96,WFGV_ISSSE_98,Tel_Eur_99}, the error rate with any planar receive array decays no faster than ${\sf SNR}^{-3}$ even though the channel is purely LoS and experiences no fading~(Section~\ref{sec:3}).

We consider the smallest number of receive antennas $n_r=4$ that can form a three-dimensional, i.e., non-planar, arrangement, and derive bounds on error performance when they form a tetrahedral array. In this case, the error probability decays faster than that of a rank~$1$ channel and is always exponential in ${\sf SNR}$ irrespective of the coding scheme used (Section~\ref{sec:3A}). 
We then design a LoS MIMO system with a good error performance for all transmit and receive array orientations over a range of inter-terminal distances by using a tetrahedral receive array and adaptively choosing two transmit antennas from a triangular/pentagonal array at the transmitter (Section~\ref{sec:system_design}).
Finally, we present simulation results to support our theoretical claims~(Section~\ref{sec:5}).

%% Notation %%
{\it Notation:} Matrices and column vectors are denoted by bold upper-case and lower-case symbols respectively. The symbols ${\bf A}^\intercal$, ${\bf A}^\dag$ and $\|{\bf A}\|_F$ denote the transpose, the conjugate-transpose and the Frobenius norm of a matrix ${\bf A}$. The symbol $\| \cdot \|$ denotes the $2$-norm of a vector.
For a complex number $z$, ${\rm arg}(z)$ and ${\rm Re}(z)$ denote its phase and real part, respectively. The expectation operator is denoted by $\mathbb{E}(\cdot)$. 

\section{The $2 \times n_r$ LoS MIMO Channel} \label{sec:2}

We consider MIMO line-of-sight (LoS) transmission with \mbox{$n_t=2$} antennas at the transmitter and \mbox{$n_r \geq 2$} antennas at the receiver. 
Assuming that the large scale fading effects, such as path loss, are accounted for in the link budget, we take the magnitude of the complex channel gain between any transmit-receive antenna pair to be unity.
If $r_{m,n}$ is the distance between the $n^{\rm th}$ transmit and the $m^{\rm th}$ receive antennas, then the $(m,n)^{\rm th}$ component of channel matrix ${\bf H} \in \mathbb{C}^{n_r \times 2}$ is~\cite{BOO_JWCOM_07}
\begin{equation} \label{eq:h_mn}
h_{m,n}=\exp\left( i\frac{2\pi r_{m,n}}{\lambda} \right),
\end{equation}
where $\lambda$ is the carrier wavelength and \mbox{$i=\sqrt{-1}$}. The resulting wireless channel is 
% \begin{equation*}
${\bf y}_{\sf Rx} = \sqrt{{\sf SNR}}{\bf Hx} + {\bf w}_{\sf Rx}$,
% \end{equation*}
where \mbox{${\bf y}_{\sf Rx} \in \mathbb{C}^{n_r}$} is the received vector, \mbox{${\bf x} \in \mathbb{C}^2$} is the transmitted vector, \mbox{${\bf w}_{\sf Rx} \in \mathbb{C}^{n_r}$} is the circularly symmetric complex white Gaussian noise with unit variance per complex dimension, and ${\sf SNR}$ is the signal-to-noise ratio at each receive antenna. The power constraint at the transmitter is \mbox{$\mathbb{E}\left(\|{\bf x}\|^2\right) \leq 1$}.
We assume that the channel matrix ${\bf H}$ is known at the receiver but not at the transmitter.
Let ${\bf h}_1,{\bf h}_2 \in \mathbb{C}^{n_r}$ denote the two columns of ${\bf H}$, and ${\bf H}={\bf QR}$ be its QR~decomposition where ${\bf Q} \in \mathbb{C}^{n_r \times 2}$ has orthonormal columns, i.e., ${\bf Q}$ is a semi-unitary matrix, and
\begin{equation*}
\renewcommand*{\arraystretch}{1.5}
{\bf R} = \begin{bmatrix} 
           \|{\bf h}_1\| & \frac{{\bf h}_1^\dag{\bf h}_2}{\| {\bf h}_1 \|} \\
            0  & \sqrt{{\|{\bf h}_2\|}^2 - \frac{{|{\bf h}_1^\dag{\bf h}_2|}^2}{{\|{\bf h}_1\|}^2}}
          \end{bmatrix}.
\end{equation*}
Let $\mu$ denote the correlation between the two columns ${\bf h}_1$ and ${\bf h}_2$ of ${\bf H}$, and $\theta_\mu$ be the phase of ${\bf h}_1^\dag{\bf h}_2$, i.e., 
\begin{align*}
\mu = \frac{|{\bf h}_1^\dag{\bf h}_2|}{\| {\bf h}_1 \| \, \|{\bf h}_2\|} \textrm{ and } \theta_\mu = {\rm arg}\left( {\bf h}_1^\dag{\bf h}_2 \right). 
\end{align*}
From~\eqref{eq:h_mn}, we have $\|{\bf h}_1\|=\|{\bf h}_2\|= \sqrt{n_r}$, and hence,
\begin{equation} \label{eq:R_original}
{\bf R} = \sqrt{n_r} \begin{bmatrix}
                      1 & {\mathrm{e}}^{i \theta_\mu} \mu \\
                      0 & \sqrt{1-\mu^2}
                     \end{bmatrix}. 
\end{equation}
Since ${\bf Q}$ is semi-unitary and ${\bf w}_{\sf Rx}$ is a white Gaussian noise vector, \mbox{${\bf y}= {\bf Q}^\dag{\bf y}_{\sf Rx}$} is a sufficient statistic for ${\bf x}$. Hence, in the rest of the paper we will consider the following equivalent channel
\begin{equation} \label{eq:effective_channel}
{\bf y} = \sqrt{{\sf SNR}}{\bf Rx} + {\bf w},
\end{equation}
where ${\bf R}$ is given in~\eqref{eq:R_original}, and \mbox{${\bf w} = {\bf Q}^\dag{\bf x}$} is a two-dimensional circularly symmetric complex white Gaussian noise with zero mean and unit variance per complex dimension.

\subsection{Modelling the ${\bf R}$ matrix}

To analyze the error performance of arbitrary coding schemes in LoS MIMO channels, we model the phase $\theta_\mu$ as independent of $\mu$ and uniformly distributed in $[0,2\pi)$. Deriving the probability distribution of $\theta_\mu$ and $\mu$ appears difficult, however, we provide an analytical motivation and numerical examples to support the validity of our model.

\begin{figure}[!t] 
\centering
\includegraphics[width=3.3in]{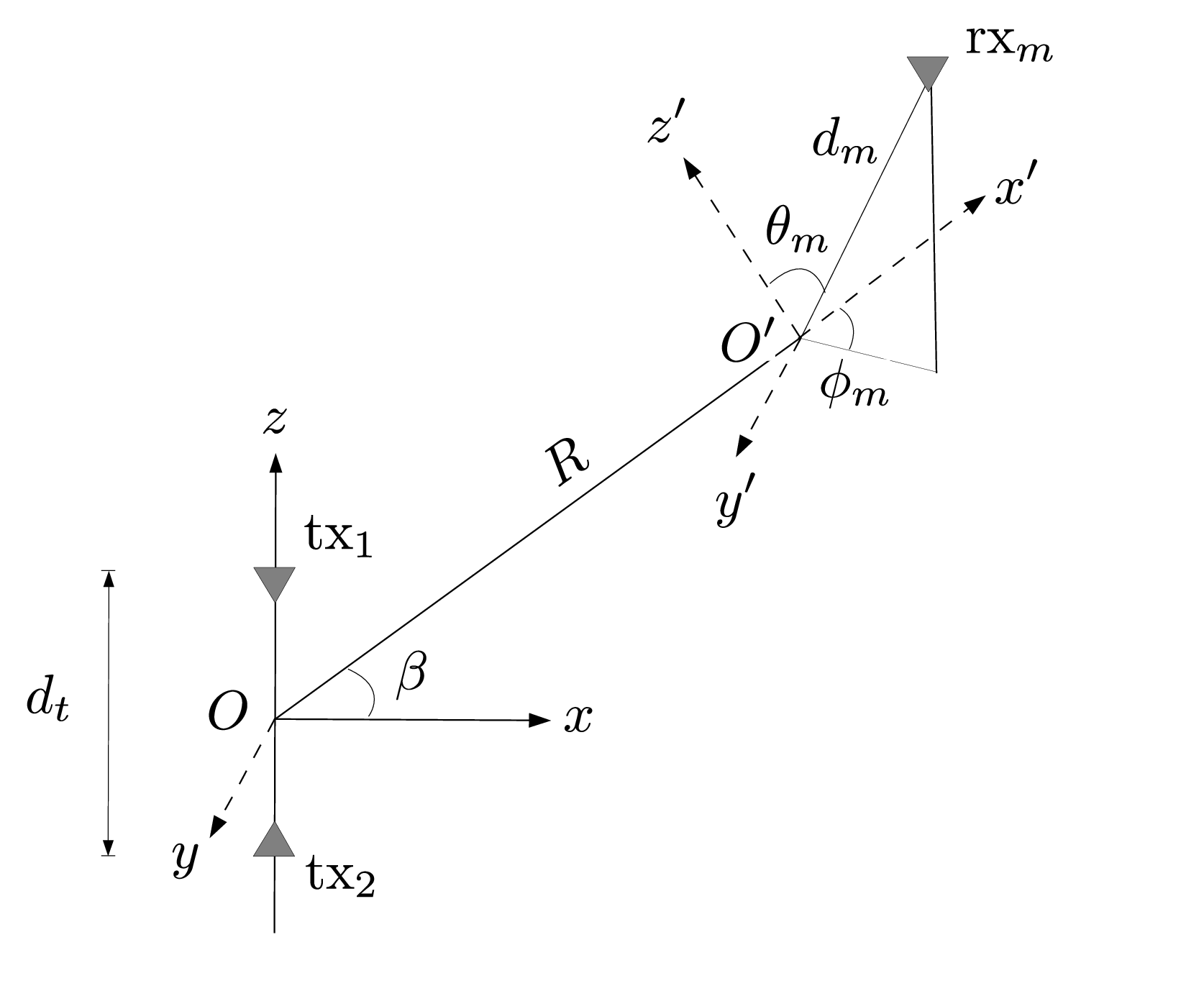}
\caption{Illustration of the parameters used in the system model.}
\label{fig:geometry_los}
\vspace{-4mm}
\end{figure}

We follow the notations from~\cite{BOO_WCNC_05,BOO_JWCOM_07} to describe the geometry of the transmit and receive antenna positions as illustrated in Fig.~\ref{fig:geometry_los}. 
%
%%%%%%%%%%%%%%%%%%%%%%%%%%%%%%%
%%%%%%%%%%%%%%%%%%%%%%%%%%%%%%
%
We denote the inter-antenna distance at the transmitter by $d_t$, and define the origin $O$ of the three-dimensional reference coordinate system as the mid-point between the two transmit antennas. 
Define the $z$-axis of the coordinate system to be along the line connecting the two transmit antennas, i.e., the positions of the two transmit antennas are ${\begin{bmatrix} 0 , \, 0 , \, \frac{d_t}{2}\end{bmatrix}}^\intercal$ and ${\begin{bmatrix} 0 , \, 0 , \,-\frac{d_t}{2}\end{bmatrix}}^\intercal$, respectively.
Choose the $x$-axis of the coordinate system such that the centroid $O'$ of the receive antenna array lies on the $x$--$z$ plane. Let $O'$ be at a distance of $R$ from $O$ and at an angle $\beta$ to the $x$-axis i.e., at the point ${\begin{bmatrix} R\cos\beta , \, 0 , \, R\sin\beta\end{bmatrix}}^\intercal$.
Consider an auxiliary coordinate system with $O'$ as the origin and the three axes $x',y',z'$ defined as follows: 
\begin{inparaenum}
\item[] the $x'$ axis is along the direction $OO'$, i.e., along the vector ${\begin{bmatrix} \cos\beta , \, 0 , \, \sin\beta \end{bmatrix}}^\intercal$, 
\item[] $z'$ axis is on the $x$--$z$ plane, and 
\item[] $y'$ is parallel to $y$. 
\end{inparaenum}
Let $\left(d_m,\theta_m,\phi_m\right)$ be the spherical coordinates of the $m^{\rm th}$ receive antenna with respect to this auxiliary coordinate system, where $d_m$ is the radial distance, $\theta_m$ is the polar angle and $\phi_m$ is the azimuthal angle.
The distance $r_{m,n}$ between the $n^{\text{th}}$ transmit and $m^{\text{th}}$ receive antennas satisfies~\cite{BOO_Eur_07}\footnote{The angle $\beta$ is equal to the parameter $\theta_t$ used in~\cite{BOO_WCNC_05,BOO_JWCOM_07}.} 
\begin{align*}
r_{m,n} &\approx R + d_m\sin\theta_m\cos\phi_m \, + (-1)^n\frac{d_t}{2}\sin\beta \, + \\ &\frac{(d_m\sin\theta_m\sin\phi_m)^2 + (d_m\cos\theta_m + (-1)^n\frac{d_t}{2}\cos\beta)^2}{2R}.
\end{align*}
Therefore, the difference $r_{m,2}-r_{m,1}$ is given by
\begin{align}
r_{m,2}-r_{m,1}&=d_t\sin\beta + \frac{(d_m\cos\theta_m + \frac{d_t}{2}\cos\beta)^2}{2R} \nonumber \\ 
&~~~~~~~~~~~~~~~~~~~~-\frac{(d_m\cos\theta_m - \frac{d_t}{2}\cos\beta)^2 }{2R} \nonumber \\
&=d_t\sin\beta + \frac{d_td_m\cos\beta\cos\theta_m}{R}. \label{eq:r_mn_dif}
\end{align}

Let \mbox{$F(\beta)={\bf h}_1^\dag{\bf h}_2$} denote the inner product between the two columns of ${\bf H}$ as a function of $\beta$. Using~\eqref{eq:h_mn} and~\eqref{eq:r_mn_dif}, we obtain
\begin{align} \label{eq:h1h2}
F(\beta)&={\bf h}_1^\dag{\bf h}_2=\sum_{m=1}^{n_r}h_{m,1}^\dag h_{m,2} \nonumber \\
        &=\exp \! \left(\frac{i2\pi d_t\sin\beta}{\lambda}\right) \sum_{m=1}^{n_r} \! \! \exp \! \left( \frac{i2\pi d_t d_m \cos\beta\cos\theta_m}{R\lambda} \right)
\end{align}
Let $f_1(\beta) = \exp\left(i2\pi d_t\sin\beta/\lambda\right)$ and
\begin{align*}
%% f_1(\beta) &= \exp\left(\frac{i2\pi d_t\sin\beta}{\lambda}\right), \text{ and} \\
f_2(\beta) &= \sum_{m=1}^{n_r}\exp \left( \frac{i2\pi d_t d_m \cos\beta\cos\theta_m}{R\lambda} \right). 
\end{align*}
Then $F(\beta)=f_1(\beta)f_2(\beta)$, $\arg F = \arg f_1 + \arg f_2$, and since $|f_1|=1$, we also have $|F|=|f_2|$.

We now upper bound the magnitude of the derivative of $\mu$ with respect to $\beta$. 
The derivative of $\mathrm{d}f_2/\mathrm{d}\beta$ equals
\begin{equation}
\sum_{m=1}^{n_r}\frac{-i2\pi d_t d_m \sin\beta\cos\theta_m}{R\lambda} \exp\left(\frac{i2\pi d_t d_m \cos\beta\cos\theta_m}{R\lambda} \right).  \label{eq:beta_rate_of_change:2}
%% \resizebox{0.9\hsize}{!}{$\frac{\mathrm{d} f_2}{\mathrm{d}\beta} =  \sum_{m=1}^{n_r}\frac{-i2\pi d_t d_m \sin\beta\cos\theta_m}{R\lambda} \exp\left(\frac{i2\pi d_t d_m \cos\beta\cos\theta_m}{R\lambda} \right).$}  \label{eq:beta_rate_of_change:2}
\end{equation} 
Note that $\left\vert {\mathrm{d} f_2}/{\mathrm{d}\beta} \right\vert \leq b$, where $\displaystyle b=\frac{2\pi d_t \sum_{m=1}^{n_r}d_m}{R\lambda}$. 
For an infinitesimal change $\Delta\beta$ in the value of $\beta$,
\begin{align*}
|f_2(\beta + \Delta\beta)| - |f_2(\beta)| = \Big\vert f_2(\beta)+\frac{\mathrm{d} f_2}{\mathrm{d}\beta} \Delta\beta \Big\vert - |f_2(\beta)|.
\end{align*} 
Using the fact that $\big\vert \, |u+w| - |u| \, \big\vert \leq |w|$ for any $u,w \in \mathbb{C}$, we have 
\begin{align*}
\Big\vert \, |f_2(\beta + \Delta\beta)| - |f_2(\beta)| \, \Big\vert \leq \left\vert \frac{\mathrm{d} f_2}{\mathrm{d}\beta} \right\vert |\Delta\beta| \leq b|\Delta\beta|.
\end{align*} 
It follows immediately that \mbox{$\left\vert \, {\mathrm{d} |f_2|}/{\mathrm{d}\beta} \, \right\vert \leq b$}.
Using the fact that \mbox{$\mu = |F(\beta)|/n_r=|f_2(\beta)|/n_r$}, we have
\begin{align} \label{eq:dmu_dbeta}
\left\vert \frac{\mathrm{d} \mu}{\mathrm{d}\beta} \right\vert = \frac{1}{n_r}\left\vert \frac{\mathrm{d} |f_2|}{\mathrm{d}\beta} \right\vert \leq  \frac{b}{n_r}.
\end{align} 

Note that $\theta_\mu = \arg F = \arg f_1 + \arg f_2$, and hence,
${\mathrm{d} \theta_\mu}/{\mathrm{d} \beta} = {\mathrm{d} (\arg f_1)}/{\mathrm{d} \beta} + {\mathrm{d} (\arg f_2)}/{\mathrm{d} \beta}$.
Now, $\arg f_1={2\pi d_t \sin \beta}/{\lambda}$, and hence, ${\mathrm{d} (\arg f_1)}/{\mathrm{d} \beta}= {2\pi d_t\cos\beta}/{\lambda}$.
Using~\eqref{eq:dmu_dbeta} and the fact that the range of transmission $R$ is much larger than $d_m$, we have 
\begin{align*}
\frac{\mathrm{d} (\arg f_1)}{\mathrm{d} \beta} = \frac{2\pi d_t\cos\beta}{\lambda} \gg \frac{2\pi d_t }{\lambda} \, \frac{\sum_{m=1}^{n_r}d_m}{R \, n_r} = \frac{b}{n_r} \geq \left\vert \frac{\mathrm{d} \mu}{\mathrm{d} \beta} \right\vert .
\end{align*} 
Hence, we expect ${\mathrm{d} \theta_\mu}/{\mathrm{d} \beta} \gg \left\vert {\mathrm{d} \mu}/{\mathrm{d} \beta} \right\vert$,
i.e., a small change in the value of $\beta$, that causes a negligible change in $\mu$, changes the phase $\theta_\mu$ by an entire cycle of $2\pi$~rad. 
This motivates the channel model where $\theta_\mu$ is independent of $\mu$ and uniformly distributed in the interval $[0,2\pi)$. 

\begin{example} \label{ex:theta_mu_variation}
Consider a \mbox{$2 \times 2$} LoS system operating in E-band at the frequency of $72$~GHz over a distance \mbox{$R=10$}~m. 
Let the two receive antennas be positioned such that $\theta_1=0$, $\theta_2=\pi$, $\phi_1=\phi_2=0$ and $d_1=d_2={d_r}/{2}$. Then, using~\eqref{eq:h1h2}, we have
\begin{align*}
{\bf h}_1^\dag{\bf h}_2 = 2 \exp\left(\frac{i2\pi d_t\sin\beta}{\lambda}\right) \cos \left( \frac{\pi d_t d_r \cos\beta}{R\lambda} \right).
\end{align*}
It follows that 
\begin{align} \label{eq:example_beta_changes}
\mu = \cos \left( \frac{\pi d_t d_r \cos\beta}{R\lambda} \right) \textrm{ and } \theta_\mu = \frac{2\pi d_t\sin\beta}{\lambda}.
\end{align}
Suppose the antenna geometry is to be configured so that ${\bf H}$ is unitary, i.e., $\mu=0$, under the assumption that $\beta=0$. This can be achieved by choosing $d_t$ and $d_r$ so that
\begin{align*}
\frac{d_t d_r \cos\beta}{R\lambda}=\frac{d_t d_r}{R\lambda} = \frac{1}{2}.
\end{align*}
This is the criterion for uniform linear arrays given in~\cite{BOO_WCNC_05,BOO_JWCOM_07,BOO_Eur_07}. With \mbox{$\lambda=4.2$}~mm, the choice of 
% \begin{equation*}
\mbox{$d_t=d_r = \sqrt{{R\lambda}/{2}} = 0.145\textrm{~m}$}
% \end{equation*}
yields \mbox{$\mu=0$}. With this choice of $d_t$ and $d_r$, through direct computation using~\eqref{eq:example_beta_changes}, we observe that as $\beta$ undergoes a small variation in value from $0$~rad through $0.029$~rad ($1.66^\circ$), the corresponding value of $\mu$ changes from $0$ to \mbox{$6.6 \times 10^{-4}$}, while $\theta_\mu$ ranges over the entire interval from $0$ to $2\pi$~rad. 
\end{example}

\begin{example} \label{ex:theta_mu_variation:2}
\begin{figure}[!t]
\centering
\includegraphics[width=3in]{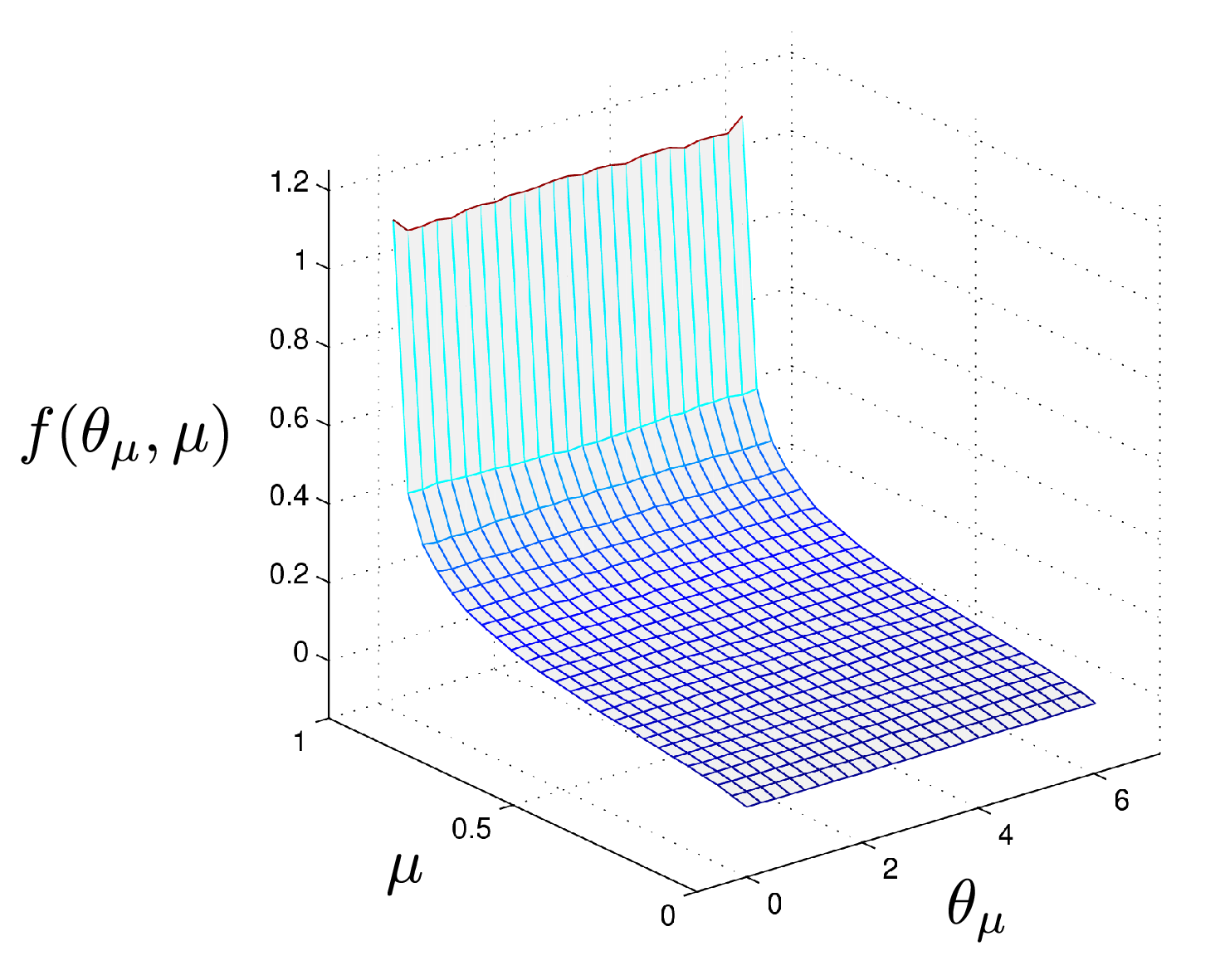}
\caption{The joint probability density function $f(\theta_\mu,\mu)$ of Example~\ref{ex:theta_mu_variation:2}.}
\label{fig:histogram}
\end{figure} 
Continuing with the $2 \times 2$ system of Example~\ref{ex:theta_mu_variation}, now assume that the transmit and receive arrays are affected by independent random rotations about their respective centroids. 
The random rotations are uniformly distributed over the space of all $3$-dimensional rotations.
The channel matrix ${\bf H}$, and the parameters $\theta_\mu$ and $\mu$ are now random variables. The joint probability density function $f(\theta_\mu,\mu)$ obtained using Monte-Carlo methods is shown in Fig.~\ref{fig:histogram}. We computed $f(\theta_\mu,\mu)$ over a rectangular grid of $625$ points using $10^7$ randomly generated instances of ${\bf H}$.
For any fixed $\mu$, we observe that $f(\theta_\mu,\mu)$ is essentially constant across all values of $\theta_\mu$, implying that $\theta_\mu$ is uniformly distributed in $[0,2\pi)$ and is independent of $\mu$.
\end{example}

\begin{example} \label{ex:theta_mu_variation:3}
Consider a $2 \times 4$ LoS MIMO system, with a rectangular array at the receiver, carrier frequency of $72$~GHz, and inter-terminal distance of $R=10$~m. 
The receive antennas are placed at the vertices of a square whose edges are of length $d_r$. We choose $d_t=d_r=\sqrt{{R\lambda}/{2}}$, which yields the ideal channel (i.e., $\mu=0$) if the transmit and receive arrays are placed broadside to each other~\cite{BOO_Eur_07}.
The joint probability density function $f(\theta_\mu,\mu)$, obtained using Monte-Carlo methods, when the transmit and receive arrays undergo uniformly random rotations about their centroids is shown in Fig.~\ref{fig:histogram_2x4}.  
As in Example~\ref{ex:theta_mu_variation:2}, the numerical result supports the validity of our channel model.
\begin{figure}[!t]
\centering
\includegraphics[width=3in]{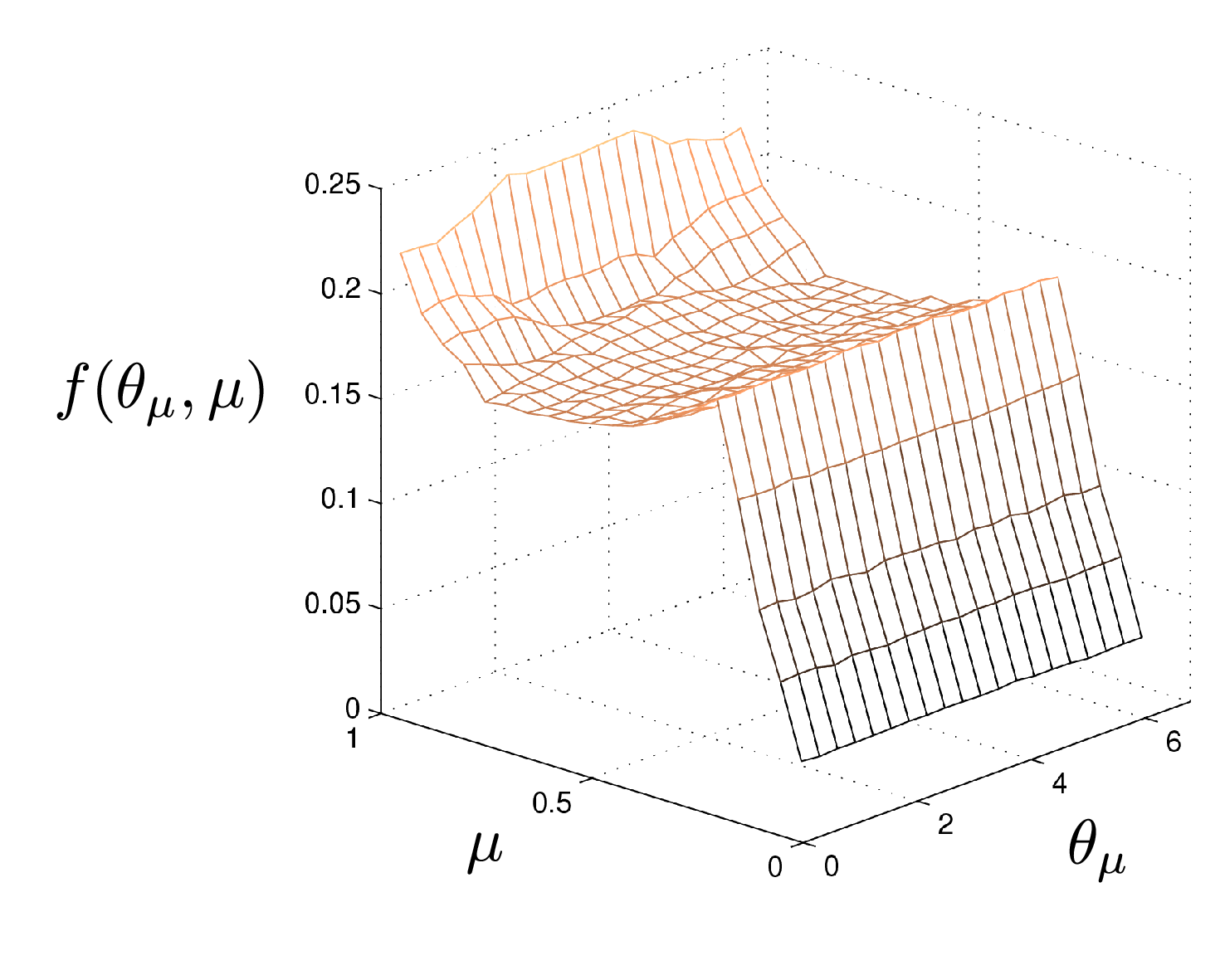}
\caption{The joint probability density function $f(\theta_\mu,\mu)$ of Example~\ref{ex:theta_mu_variation:3}.}
\label{fig:histogram_2x4}
\end{figure} 
\end{example}

In the rest of the paper we model the $2 \times n_r$ LoS channel using the $2 \times 2$ matrix (cf.~\eqref{eq:effective_channel})
\begin{equation} \label{eq:R_model}
{\bf R} = \sqrt{n_r} \begin{bmatrix} 1 & {\mathrm{e}}^{i\Theta} \mu \\ 0 & \sqrt{1-\mu^2} \end{bmatrix},
\end{equation}
where $\Theta$ is uniformly distributed in $[0,2\pi)$ and 
\begin{equation} \label{eq:mu_model}
\mu = \frac{1}{n_r} \left\vert \sum_{m=1}^{n_r}\exp \left( \frac{i2\pi d_t d_m \cos\beta\cos\theta_m}{R\lambda} \right) \right\vert.
\end{equation} 

% % % Altered for R1 version % % % % 
\subsection{Coding schemes}

We analyse the error performance of any arbitrary coding scheme for two transmit antennas with a finite transmission duration.
Let \mbox{$T \geq 1$} denote the transmission duration of a given communication scheme and \mbox{$\mathscr{C} \subset \mathbb{C}^{2 \times T}$} the finite set of all possible transmit codewords. The rows of the codewords \mbox{${\bf X} \in \mathscr{C}$} correspond to the two transmit antennas and the columns to the $T$ time slots. 
All codewords are equally likely to be transmitted and the optimal decoder, i.e., the maximum-likelihood (ML) decoder, is used at the receiver.
We further assume that the communication scheme satisfies the average power constraint $\sum_{{\bf X} \in \mathscr{C}} \|{\bf X}\|_F^2 \leq |\mathscr{C}|\,T$. Our analysis holds for arbitrary codes $\mathscr{C}$, including space-time block codes (STBCs)~\cite{TSC_JIT_98}. 

We now briefly recall two specific coding schemes which will be used in our simulations (in Section~\ref{sec:5}) to illustrate our analytical results.
\emph{Spatial multiplexing (SM)}~\cite{Fos_Bell_96,WFGV_ISSSE_98,Tel_Eur_99}, which is also known as \emph{VBLAST} in the literature, is a simple yet powerful scheme where independent information symbols are transmitted across different antennas and time slots. The codebook $\mathscr{C} \subset \mathbb{C}^{2 \times 1}$ corresponding to SM occupies \mbox{$T=1$} time slot, and is given by
\begin{equation*}
 \mathscr{C} = \left\{ \begin{bmatrix} s_1 \\ s_2 \end{bmatrix} ~ \Big\vert ~ s_1, s_2 \in \mathcal{A} \right\},
\end{equation*} 
where $\mathcal{A}$ is a complex constellation, such as QAM or PSK.

The \emph{Golden code}~\cite{BRV_JIT_05} is an STBC for two transmit antennas occupying \mbox{$T=2$} time slots, and is given by
\begin{equation*}
 \mathscr{C} = \left\{ \begin{bmatrix} ~\alpha (s_1 + \tau s_3) & \alpha(s_2 + \tau s_4) \\ i \bar{\alpha}(s_2 + \mu s_4) & \bar{\alpha}(s_1 + \mu s_3) \end{bmatrix} ~\Big\vert~ s_1,\dots,s_4 \in \mathcal{A} \right\},
\end{equation*} 
where $\mathcal{A}$ is a QAM constellation, $\tau={(1+\sqrt{5})}/{2}$, $\mu={1}/{\tau}$, $\alpha=1+i\mu$ and $\bar{\alpha}=1+i\tau$. 
Unlike SM, the Golden code spreads the information symbols across time and antennas. 

Both SM and Golden code have been well studied in the case of non line-of-sight MIMO fading channels. The SM scheme provides high data rate with low complexity encoding and decoding, while the Golden code provides high data rate, full-diversity as well as a large coding gain at the cost of higher decoding complexity in fading channels.

% % % % % % % % % % % % % % % % 

\subsection{Error probability analysis for a fixed $\mu$} \label{sec:2B}

% % Altered for Revision 1 % % 
We now analyse the error performance of a given arbitrary coding scheme for a fixed value of $\mu$. 
Let $\mathscr{C} \subset \mathbb{C}^{2 \times T}$ be any code and ${\bf X}_a,{\bf X}_b \in \mathscr{C}$ be two distinct codewords. Let \mbox{$\Delta{\bf X} = {\bf X}_a - {\bf X}_b$} be the pairwise codeword difference matrix.
% % % % % % % % % % % % % % % % % % % %
The pairwise error probability between ${\bf X}_a$ and ${\bf X}_b$ for a fixed $\mu$ and a given realization $\Theta=\theta$ is~\cite{TSC_JIT_98} 
\begin{align*}
{\sf PEP}\left( {\bf X}_a \to {\bf X}_b | \mu, \Theta=\theta \right) = \mathcal{Q}\left( \sqrt{ \frac{{\sf SNR}\|{\bf R}\Delta{\bf X}\|_F^2}{2} }\right),
\end{align*}
where $\mathcal{Q}$ is the Gaussian tail function. Using the Chernoff bound $\displaystyle \mathcal{Q}(x) \leq \frac{\exp\left( -{x^2}/{2} \right)}{2}$, we have the upper bound
\begin{equation} \label{eq:PEP_basic}
{\sf PEP} \leq \frac{1}{2} \exp \left( -\frac{{\sf SNR}}{4}\|{\bf R}\Delta{\bf X}\|_F^2 \right).
\end{equation}
Denoting the two rows of the matrix $\Delta{\bf X}$ as $\Delta{\bf x}_1^\intercal$ and $\Delta{\bf x}_2^\intercal$, 
we obtain the following expression for the squared Euclidean distance between the codewords at the receiver,
\begin{align}
\|{\bf R}\Delta{\bf X}\|_F^2 & = n_r \! \left( \| \Delta{\bf x}_1 \|^2 + \| \Delta{\bf x}_2 \|^2 + 2 \mu \, {\rm Re}({\mathrm{e}}^{i\theta}\Delta{\bf x}_1^\dag\Delta{\bf x}_2) \right) \nonumber  \\
                             & = n_r \! \left( \| \Delta{\bf x}_1 \|^2 + \| \Delta{\bf x}_2 \|^2 + 2 \mu \cos\theta' \vert\Delta{\bf x}_1^\dag\Delta{\bf x}_2\vert \right) \label{eq:distance_theta}
\end{align}
where \mbox{$\theta' = \theta + {\rm arg}(\Delta{\bf x}_1^\dag\Delta{\bf x}_2)~{\rm mod~} 2\pi$}.  

\subsubsection{Worst-case Error Probability over $\theta$}

For a given $\mu$, the value of $\theta$ that minimizes the squared Euclidean distance $\|{\bf R}\Delta{\bf X}\|^2$ at the receiver is \mbox{$\theta^*=\pi + {\rm arg}(\Delta{\bf x}_1^\dag\Delta{\bf x}_2)$} since it leads to \mbox{$\cos\theta'=-1$} in~\eqref{eq:distance_theta}.
Using the notation
\begin{align} \label{eq:d_first}
{\sf d}(\mu,\Delta{\bf X}) = \| \Delta{\bf x}_1\|^2 + \|\Delta{\bf x}_2\|^2 - 2\mu\vert \Delta{\bf x}_1^\dag\Delta{\bf x}_2 \vert,
\end{align} 
the worst-case squared Euclidean distance is
\begin{equation*}
% n_r\| \Delta{\bf x}_1 \|^2 + n_r\| \Delta{\bf x}_2 \|^2 - 2n_r\mu \vert\Delta{\bf x}_1^\dag\Delta{\bf x}_2\vert 
\min_{\theta \in [0,2\pi)}\|{\bf R}\Delta{\bf X}\|_F^2 = n_r{\sf d}(\mu,\Delta{\bf X}).
\end{equation*} 
Thus the worst-case ${\sf PEP}$ for a fixed $\mu$ satisfies
\begin{align} \label{eq:PEP_worst_theta}
{\sf PEP}^*(\mu) \leq \frac{1}{2} \exp \left( \frac{-n_r \, {\sf SNR} \, {\sf d}(\mu,\Delta{\bf X})}{4} \right).
\end{align}

\subsubsection{Average Error Probability over $\Theta$} \label{sec:2B2}

Since $\Theta$ is uniformly distributed in $[0,2\pi)$, so is $\Theta'=\Theta + {\rm arg}(\Delta{\bf x}_1^\dag\Delta{\bf x}_2)~{\rm mod~} 2\pi$.
Using~\eqref{eq:PEP_basic} and~\eqref{eq:distance_theta}, the error probability averaged over $\Theta$, for a fixed $\mu$, can be upper bounded as follows
\begin{align*}
\mathbb{E}_{\Theta}\left( {\sf PEP} \right) &\leq \mathbb{E}_{\Theta} \left( \frac{1}{2} \exp\left( -\frac{{\sf SNR}}{4} \| {\bf R}\Delta{\bf X} \|_F^2 \right) \right) \\
&= \frac{1}{2}\exp\left(\frac{-{\sf SNR}n_r(\| \Delta{\bf x}_1 \|^2 + \| \Delta{\bf x}_2 \|^2)}{4}\right) \, \times \\ 
&~~~\frac{1}{2\pi} \int_0^{2\pi} \! \! \! \! \exp\left( -\frac{{\sf SNR}n_r}{4} 2\mu \cos\theta' \vert\Delta{\bf x}_1^\dag\Delta{\bf x}_2\vert\right) \, \mathrm{d}\theta' \\
& = \frac{1}{2}\exp\left(\frac{-{\sf SNR}n_r(\| \Delta{\bf x}_1 \|^2 + \| \Delta{\bf x}_2 \|^2)}{4}\right) \, \times \\ 
&~~~~~~~~~~~~~~~~~~~~~~~~~~~~~I_0\left( \frac{{\sf SNR}n_r}{2} \mu \vert\Delta{\bf x}_1^\dag\Delta{\bf x}_2\vert \right)
\end{align*}
where
\begin{align*}
I_0(x) &= \frac{1}{\pi} \int_0^{\pi} \! \exp\left( x \cos\theta' \right) \, \mathrm{d}\theta' = \frac{1}{2\pi} \int_0^{2\pi} \! \exp\left( x \cos\theta' \right) \, \mathrm{d}\theta' \\
       &= \frac{1}{2\pi} \int_0^{2\pi} \! \exp\left( -x \cos\theta' \right) \, \mathrm{d}\theta'
\end{align*}
is the modified Bessel function of the first kind and zeroth order. For large $x$ we have~\cite{AbS_NBS_64}
\begin{align} \label{eq:bessel_approx}
I_0(x) = \frac{{\mathrm{e}}^{x}}{\sqrt{2\pi x}} \left( 1 + O\left( {x}^{-1} \right) \right).
\end{align}
Using~\eqref{eq:d_first} 
and the first order approximation~\eqref{eq:bessel_approx}, we get the following approximate upper bound when $\mu > 0$,
\begin{align} \label{eq:PEP_avg_theta}
\mathbb{E}_\Theta \left( {\sf PEP} \right) &\lesssim \frac{1}{\sqrt{4 \pi n_r {\sf SNR} \mu \vert\Delta{\bf x}_1^\dag\Delta{\bf x}_2\vert}} \nonumber \\
&~~~~~~~~~~~\times \exp\left( -\frac{n_r {\sf SNR}}{4} {\sf d} \! \left( \mu, \Delta{\bf X} \right)\right).
\end{align}
Since the exponential function falls more rapidly than ${\sf SNR}^{-{1}/{2}}$, the high ${\sf SNR}$ behaviour is dictated by ${\sf d}(\mu,\Delta{\bf X})$. 

In this section, we derived bounds on ${\sf PEP}$ for a fixed $\mu$.
In Sections~\ref{sec:3} and~\ref{sec:4} we analyze the effects of random rotations of the terminals on $\mu$ and error performance.

\section{Error performance of planar receive arrays} \label{sec:3} 

Assume that the receive antenna system is affected by a random three-dimensional rotation \mbox{${\bf U} \in \mathbb{R}^{3 \times 3}$} about its centroid $O'$. Let the rotation ${\bf U}$ be uniformly distributed on the set of all $3$-dimensional rotations, i.e., the special orthogonal group 
\begin{equation*}
SO_3=\left\{ {\bf U} \in \mathbb{R}^{3 \times 3}~|~{\bf UU}^\intercal = {\bf I}, \det({\bf U})=1 \right\}.
\end{equation*}
In Theorem~\ref{thm:planar_lower_bound}, we provide a lower bound on the average pairwise error probability over a LoS MIMO channel with planar receive array. To do so, we derive a lower bound on the probability that a random rotation ${\bf U}$ would lead to a `bad' channel matrix with $\mu$ close to $1$, i.e. $\mu \geq 1 -\epsilon$ for some small positive $\epsilon$. By analyzing the ${\sf PEP}$ for this class of bad channels, and letting $\epsilon$ decay suitably with ${\sf SNR}$, we arrive at a lower bound for the average ${\sf PEP}$ at high ${\sf SNR}$.

\begin{theorem} \label{thm:planar_lower_bound}
Let the receive antenna array be any planar arrangement of $n_r$ antennas, $n_r \geq 2$, undergoing a uniformly distributed random rotation ${\bf U}$ about its centroid. At high ${\sf SNR}$, for any transmit orientation $\beta$, we have
\begin{align} \label{eq:PEP_avg_planar}
\mathbb{E}({\sf PEP}) &\geq \frac{\exp \left( -\frac{n_r {\rm c} \, |\Delta{\bf x}_1^\dag \Delta{\bf x}_2| }{2} \right)   }{2n_r{\sf SNR}^3  \sqrt{2\pi^2 |\Delta{\bf x}_1^\dag \Delta{\bf x}_2|}   \left( \| \Delta{\bf X} \|_F + \frac{1}{\sqrt{n_r{\sf SNR}}} \right)} \nonumber \\
&~~~~~~~~~~~ \times \exp \left( -\frac{n_r{\sf SNR}}{4} \, {\sf d}(1,\Delta{\bf X}) \right),
\end{align}
where ${\rm c} = \max_{m=1}^{n_r} {2\pi d_t d_m}/{R\lambda}$.
\end{theorem}
\begin{IEEEproof}
Let $\{{\bf e}_x,{\bf e}_y,{\bf e}_z\}$ be the standard basis in $\mathbb{R}^3$. When the receive system undergoes no rotation, i.e., when ${\bf U}={\bf I}$, let the position of the $m^{th}$ receive antenna relative to the centroid $O'$ of the receive antenna system be $d_m{\bf r}_m$, where ${\bf r}_m \in \mathbb{R}^3$ is a unit vector. 
Since the receive array is planar and the random rotation ${\bf U}$ is uniformly distributed, without loss of generality, we assume that the vectors ${\bf r}_1,\dots,{\bf r}_{n_r}$ are in the linear span of ${\bf e}_x$ and ${\bf e}_z$.
From Fig.~\ref{fig:geometry_los} we see that $\theta_m$ in~\eqref{eq:h1h2} is the angle between the orientation ${\bf Ur}_m$ of the $m^{th}$ receiver and the unit vector ${\bf \tilde{v}}={\begin{bmatrix} -\sin\beta , \, 0 , \, \cos\beta \end{bmatrix}}^\intercal$ along $z'$-axis, i.e., $\cos\theta_m = {\bf r}_m^\intercal{{\bf U}}^\intercal{\bf \tilde{v}}$. 
Note that ${\bf U}^\intercal$ has the same distribution as ${\bf U}$, and ${\bf v}={\bf U}^\intercal{\bf \tilde{v}}$ is uniformly distributed on the unit sphere in $\mathbb{R}^3$. The resulting random variable $| {\bf e}_y^\intercal {\bf v}|$ is known to be uniformly distributed in the interval $[0,1]$. 

For a small positive number $\delta > 0$, consider the event \mbox{$\mathcal{E}:|{\bf e}_y^\intercal{\bf v}|^2 \geq 1 - \delta^2$}. The probability of $\mathcal{E}$ is 
\begin{align*}
{\sf P}(\mathcal{E}) = {\sf P}\left(|{\bf e}_y^\intercal{\bf v}| \geq \sqrt{1-\delta^2}\right) = 1 - \sqrt{1-\delta^2} \approx \frac{\delta^2}{2},
\end{align*}
for small values of $\delta$. We will now derive an upper bound for the ${\sf PEP}$ for the case when $\mathcal{E}$ is true. Using the following inequalities, we first show that \mbox{$|\cos\theta_m| \leq \delta$}, for all $m=1,\dots,n_r$,
\begin{align*}
{|\cos\theta_m|}^2 &= |{\bf r}_m^\intercal{\bf v}|^2 \\
                   &\leq |{\bf e}_x^\intercal{\bf v}|^2 + |{\bf e}_z^\intercal{\bf v}|^2 \hfill \textrm{ (since } {\bf r}_m \in {\rm span}({\bf e}_x,{\bf e}_z) {\rm)} \\
                 &= {\| {\bf v} \|}^2 - |{\bf e}_y^\intercal{\bf v}|^2 \\
                 &\leq 1 - (1-\delta^2) = \delta^2.
\end{align*}
Let \mbox{$c_m={2\pi d_t d_m \cos\beta}/{R\lambda}$} and \mbox{$c_{\rm max} = \max\{c_1,\dots,c_{n_r}\}$}. From~\eqref{eq:mu_model}, we have
\begin{align*}
\mu = \frac{1}{n_r} \left\vert \sum_{m=1}^{n_r} \exp\left( i c_m \cos\theta_m \right) \right\vert.
\end{align*}
We will now show that the value of $\mu$ is close to $1$ when $\mathcal{E}$ is true. If $\epsilon_m = 1 - \exp(i c_m \cos\beta)$, then
\begin{align*}
|\epsilon_m|^2 &= \left(1 - \cos(c_m\cos\theta_m) \right)^2 + \sin^2\left(c_m\cos\theta_m\right) \\
               & = 2 - 2 \cos (c_m\cos\theta_m) \\
               &\approx 2 - 2 \left( 1 - \frac{c_{m}^2\cos^2 (\theta_m)}{2} \right)\\
               & = c_m^2\cos^2(\theta_m)
               \leq \delta^2 c_{\rm max}^2,
\end{align*}
where the approximation follows from the Taylor's series expansion of the $\cos(\cdot)$ function and the fact that \mbox{$|c_m\cos\theta_m| \leq c_m\delta$} is small. Now,
\begin{align*}
\mu &= \frac{1}{n_r} \left\vert \sum_{1}^{n_r} \left(1 - \epsilon_m \right) \right\vert 
     = \frac{1}{n_r} \left\vert n_r - \sum_{1}^{n_r} \epsilon_m \right\vert \\
    &\geq 1 - \frac{1}{n_r}\sum_{1}^{n_r}|\epsilon_m| \geq 1 - \delta c_{\rm max}.
\end{align*}
Thus $\mu \geq 1 - \delta c_{\rm max}$ whenever $\mathcal{E}$ is true.

The pairwise error probability for fixed $\mu$ and $\Theta=\theta$ is $\mathcal{Q}\left( \sqrt{ {{\sf SNR}\|{\bf R}\Delta{\bf X}\|_F^2}/{2} } \right)$. Since we need a lower bound on the probability of error, we use the following lower bound for the Gaussian tail function~\cite{Bar_JMAA_08}
\begin{align*}
\mathcal{Q}(x) \geq \frac{2}{\sqrt{2\pi}\left(x+\sqrt{x^2+4}\right)}\,{\exp\left( -\frac{x^2}{2} \right)}, \textrm{ for } x \geq 0.
\end{align*}
Using $x^2+4 \leq (x+2)^2$ for $x \geq 0$, we obtain a more relaxed bound
\begin{equation*}
\mathcal{Q}(x) \geq \frac{1}{\sqrt{2\pi}(x+1)}\,{\exp\left( -\frac{x^2}{2} \right)}.
\end{equation*}
In our case $x=\sqrt{{{\sf SNR}\|{\bf R}\Delta{\bf X}\|_F^2}/{2}}$, and we use the exact value of $x$ from~\eqref{eq:distance_theta} for the exponent, and the following upper bound for the denominator
\begin{align*}
x = \sqrt{\frac{\sf SNR}{2}} \| {\bf R}\Delta{\bf X}\|_F &\leq \sqrt{\frac{\sf SNR}{2}} \, \| {\bf R} \|_F \, \, \|\Delta{\bf X}\|_F\\
                                                         &= \sqrt{n_r{\sf SNR}} \| \Delta{\bf X} \|_F.
\end{align*} 
Thus, we have the following lower bound for a fixed $\mu$ and $\Theta=\theta$,
\begin{align} \label{eq:1:thm:planar_lower_bound}
{\sf PEP} \geq \frac{\exp\left(-\frac{\sf SNR}{4} \, \|{\bf R}\Delta{\bf X}\|_F^2\right)}{\sqrt{2\pi}\left( \sqrt{n_r{\sf SNR}}\|\Delta{\bf X}\|_F + 1 \right)} .
\end{align}
Since the denominator is independent of the phase $\Theta$, we can use the same method as in Section~\ref{sec:2B2} to obtain the average of the above lower bound over the uniformly distributed random variable $\Theta$. Averaging~\eqref{eq:1:thm:planar_lower_bound} over $\Theta$ and using the approximation to the Bessel function~\eqref{eq:bessel_approx}, we obtain 
\begin{align*}
\mathbb{E}_{\Theta} \! \left({\sf PEP}\right) \gtrsim \frac{\exp \left( -\frac{n_r{\sf SNR}}{4} {\sf d}(\mu,\Delta{\bf X}) \right)}{n_r{\sf SNR}  \sqrt{2\pi^2 \mu |\Delta{\bf x}_1^\dag \Delta{\bf x}_2|}   \left( \| \Delta{\bf X} \|_F + \frac{1}{\sqrt{n_r{\sf SNR}}} \right)}
\end{align*}
Using the trivial upper bound $\mu \leq 1$ in the denominator,
\begin{align} \label{eq:proof_planar_1}
\mathbb{E}_{\Theta}\left({\sf PEP}\right) \gtrsim \frac{\exp \left( -\frac{n_r{\sf SNR}}{4} {\sf d}(\mu,\Delta{\bf X}) \right)}{n_r{\sf SNR}  \sqrt{2\pi^2 |\Delta{\bf x}_1^\dag \Delta{\bf x}_2|}   \left( \| \Delta{\bf X} \|_F + \frac{1}{\sqrt{n_r{\sf SNR}}} \right)}.
\end{align}
Since ${\sf d}(\mu,\Delta{\bf X})$ is a decreasing function of $\mu$, if $\mathcal{E}$ is true, the numerator in the RHS of~\eqref{eq:proof_planar_1} can be lower bounded by $\exp\left( -\frac{n_r{\sf SNR}}{4} {\sf d}(1-\delta c_{\rm max},\Delta{\bf X}) \right)$.
The expression~\eqref{eq:proof_planar_1} is a lower bound on the average ${\sf PEP}$ for a given $\mu$. We now derive a lower bound for the ${\sf PEP}$ when averaged over both $\mu$ and $\Theta$ as follows
\begin{align}
\mathbb{E}({\sf PEP}) &= {\sf P}(\mathcal{E}) {\sf P}\left( {\bf X}_a \to {\bf X}_b | \mathcal{E}\right) + {\sf P}(\mathcal{E}^c) {\sf P}\left( {\bf X}_a \to {\bf X}_b | \mathcal{E}^c\right) \nonumber \\
   &\geq {\sf P}(\mathcal{E}) {\sf P}\left( {\bf X}_a \to {\bf X}_b | \mathcal{E}\right) \nonumber \\
   &\geq \frac{\delta^2 \, \exp \left( -\frac{n_r{\sf SNR}}{4} {\sf d}(1-\delta c_{\rm max},\Delta{\bf X}) \right)}{2n_r{\sf SNR}  \sqrt{2\pi^2 |\Delta{\bf x}_1^\dag \Delta{\bf x}_2|}   \left( \| \Delta{\bf X} \|_F + \frac{1}{\sqrt{n_r{\sf SNR}}} \right)}. \label{eq:app_PEP_planar_lower_bound}
\end{align} 
From the definition~\eqref{eq:d_first} of ${\sf d}(\mu,\Delta {\bf X})$, we have
\begin{align*} %% \label{eq:app_d_large_mu}
{\sf d}(1-\delta c_{\rm max},\Delta{\bf X})= {\sf d}(1,\Delta{\bf X}) + 2\delta c_{\rm max} |\Delta{\bf x}_1^\dag\Delta{\bf x}_2|,
\end{align*}
Using the above relation and choosing $\delta = {\sf SNR}^{-1}$, which is small for high ${\sf SNR}$, we obtain
\begin{align*}
 \mathbb{E}({\sf PEP}) \geq \frac{\exp \left( -\frac{n_r c_{\rm max}|\Delta{\bf x}_1^\dag\Delta{\bf x}_2| }{2} \right) \exp \left( -\frac{n_r{\sf SNR}}{4} {\sf d}(1,\Delta{\bf X}) \right)  }{2n_r{\sf SNR}^3  \sqrt{2\pi^2 |\Delta{\bf x}_1^\dag \Delta{\bf x}_2|}   \left( \| \Delta{\bf X} \|_F + \frac{1}{\sqrt{n_r{\sf SNR}}} \right)}.
\end{align*}
Using \mbox{$\cos\beta \leq 1$} in \mbox{$c_{m} = {2\pi d_t d_m \cos\beta}/{R \lambda}$} we obtain $c_{\max} \geq \max_{m} {2 \pi d_t d_m}/{R_{\min}\lambda}$. This completes the proof.
\end{IEEEproof}

We compare the lower bound~\eqref{eq:PEP_avg_planar} on ${\sf PEP}$ for planar receive arrays undergoing random rotations, with the upper bound~\eqref{eq:PEP_avg_theta} for a channel with fixed $\mu=1$. The dominant term dictating the rate of decay of error probability for both these channels is $\exp\left( -\frac{n_r \, {\sf SNR}}{4} \min_{\Delta {\bf X}} {\sf d}(1,\Delta{\bf X}) \right)$, where the minimization is over all non-zero codewords difference matrices \mbox{$\Delta{\bf X}={\bf X}_a - {\bf X}_b$} of the code $\mathscr{C}$. Note that $\mu=1$ minimizes the performance metric ${\sf d}(\mu,\Delta{\bf X})$, and corresponds to the worst-case scenario where both ${\bf H}$ and ${\bf R}$ have rank $1$. While planar receive arrays, such as the well-studied linear, rectangular and circular arrays, provide an array gain (an $n_r$-fold increase in received ${\sf SNR}$), their \emph{asymptotic coding gain} $\min_{\Delta{\bf X}} {\sf d}(1,\Delta{\bf X})$ provides no improvement over that of any rank $1$ channel.

Theorem~\ref{thm:planar_lower_bound} further implies that when \mbox{$\min_{\Delta{\bf X}}{\sf d}(1,\Delta{\bf X})=0$}, the error probability is no more exponential in ${\sf SNR}$, but decays at the most as fast as ${\sf SNR}^{-3}$. Hence, although the channel is purely LoS and experiences no fading, the error performance with a planar arrangement of antennas can decay slowly, similar to a fading channel. 

The parameter ${\sf d}(1,\Delta{\bf X})$ satisfies the following tight inequality
\begin{align}
{\sf d}(1,\Delta{\bf X}) &= \|\Delta{\bf x}_1\|^2 + \|\Delta{\bf x}_2\|^2 - 2|\Delta{\bf x}_1^\dag\Delta{\bf x}_2| \nonumber \\
                         &\geq \|\Delta{\bf x}_1\|^2 + \|\Delta{\bf x}_2\|^2 - 2\|\Delta{\bf x}_1\| \, \|\Delta{\bf x}_2\| \nonumber \\
                         &= {\left( \, \|\Delta{\bf x}_1\| -  \|\Delta{\bf x}_2\| \, \right)}^2. \label{eq:cauchy_schwarz}
\end{align}
The second line follows from the Cauchy-Schwarz inequality which is tight if and only if $\Delta{\bf x}_1$ and $\Delta{\bf x}_2$ are linearly dependent. Thus, ${\sf d}(1,\Delta{\bf X})=0$ if and only if $\Delta{\bf x}_1$ and $\Delta{\bf x}_2$ are linearly dependent and $\|\Delta{\bf x}_1\| = \|\Delta{\bf x}_2\|$, i.e., if and only if \mbox{$\Delta{\bf x}_1 = \alpha \Delta{\bf x}_2$} for some complex number $\alpha$ of unit magnitude.
We use this observation in Example~\ref{ex:vblast} below to show that the widely used spatial multiplexing coding scheme suffers from such a slowly decaying error probability with planar receive arrays.

\begin{example} \label{ex:vblast}
\emph{Performance of Spatial Multiplexing with Planar Receive Array.}
The codeword difference matrices of the SM scheme are of the form 
\begin{equation*}
 \Delta{\bf X} = \begin{bmatrix} \Delta s_1 \\ \Delta s_2 \end{bmatrix},
\end{equation*} 
where $\Delta s_1,\Delta s_2 \in \Delta \mathcal{A}$ and $\Delta \mathcal{A} = \left\{ x - y \, \vert \, x,y \in \mathcal{A} \right\}$ is the set of pairwise differences of the complex constellation $\mathcal{A}$. When $\Delta s_1 = \Delta s_2$ the two rows of the codeword difference matrix $\Delta {\bf X}$ are equal resulting in ${\sf d}(1,\Delta {\bf X})=0$.
Hence, for the SM scheme, $\min_{\Delta {\bf X}} {\sf d}(1,\Delta {\bf X})=0$, and from Theorem~\ref{thm:planar_lower_bound}, the rate of decay of the average error probability will be no faster than ${\sf SNR}^{-3}$.
Note that this result is valid for any number of antennas $n_r$ used in any planar arrangement of the receive array.
This theoretical result is validated by our simulations (see Fig.~\ref{fig:manuscript1} and Fig.~\ref{fig:manuscript3}) in Section~\ref{sec:5}.
\end{example}

\section{Error Performance of Tetrahedral Receive array} \label{sec:4}

\begin{figure}[!t] 
\centering
\includegraphics[width=2.5in]{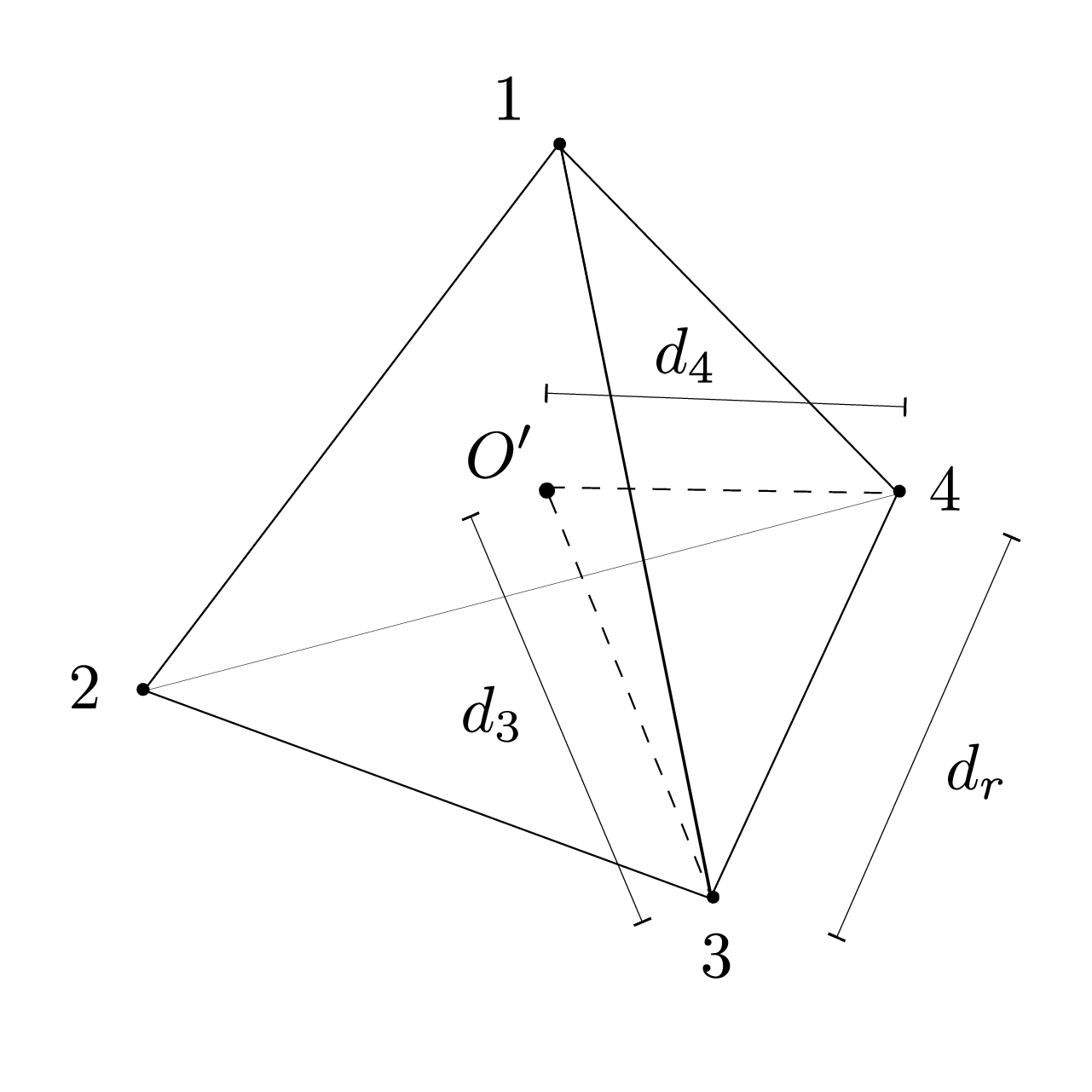}
\vspace{-3mm}
\caption{The receive antennas are placed at the vertices $1,\dots,4$ of the tetrahedron. Also shown in the figure are the centroid $O'$, the distances $d_3$ and $d_4$ of the antennas $3$ and $4$ from $O'$, and the inter-antenna distance $d_r$.}
\label{fig:tetrahedron_main}
\end{figure}

The smallest number of antennas that can form a non-planar arrangement is $4$. In this section we consider the case where $n_r=4$ receive antennas are placed at the vertices of a regular tetrahedron, see Fig.~\ref{fig:tetrahedron_main}.
The inter-antenna distance $d_r$ is the same for any pair of receive antennas, and this is related to the distance $d_m$ of each antenna from the centroid $O'$ of the receive array as \mbox{$d_m = \sqrt{{3}/{8}} \, d_r$}, $m=1,\dots,4$.
Let us define the \emph{deviation factor} $\eta$ as in~\cite{BOO_WCNC_05,BOO_JWCOM_07} as follows
\begin{align} \label{eq:eta}
\eta = \frac{R \, \lambda}{2 d_t d_r \cos\beta}.
\end{align}
In the case of a tetrahedral receiver, using~\eqref{eq:mu_model} and~\eqref{eq:eta}, 
\begin{align*}
% \textstyle
{ \mu = \frac{1}{4} \left\vert \sum_{m=1}^{4} \exp\left( i \frac{\pi}{\eta} \sqrt{\frac{3}{8}} \cos\theta_m \right) \right\vert.}
\end{align*} 
The parameter $\eta$ captures both the distance $R$ and the transmit orientation $\beta$, while the variables $\theta_1,\dots,\theta_4$ jointly determine the receive orientation ${\bf U}$. In order to upper bound the error probability using~\eqref{eq:PEP_worst_theta}, we need the maximum value of $\mu$ over all possible $\eta$ and ${\bf U}$. Let 
\begin{align} \label{eq:mu_star}
 \mu^*(\eta) = \max_{{\bf U} \in SO_3} \frac{1}{4} \left\vert \sum_{m=1}^{4} \exp\left( i \frac{\pi}{\eta} \sqrt{\frac{3}{8}} \cos\theta_m \right) \right\vert
\end{align} 
be the maximum channel correlation over all receive orientations as a function of $\eta$. If one is aware of the range of values that $R$ and $\beta$ may assume, then one can upper bound the worst-case ${\sf PEP}$ using~\eqref{eq:PEP_worst_theta} as
\begin{align} \label{eq:PEP_eta}
 {\sf PEP}^* &\leq \frac{1}{2} \exp\left( -\frac{n_r \, {\sf SNR}}{4} \, {\sf d}(\max_{\eta} \mu^*(\eta),\Delta{\bf X}) \right) \nonumber \\
             &= \frac{1}{2} \exp\left( -{\sf SNR} \, \, {\sf d}(\max_{\eta} \mu^*(\eta),\Delta{\bf X}) \right).
\end{align} 

\subsection{An upper bound on $\mu^*(\eta)$} \label{sec:3A}

In this sub-section we derive an upper bound on $\mu^*(\eta)$ for all $\eta \geq 1$. This result will allow us to show that the high ${\sf SNR}$ error performance of the tetrahedral array is better than any planar receive array when $\eta \geq 1$ and the receiver undergoes a uniformly random rotation.
To derive this upper bound, we first show that when $\eta \geq 1$, irrespective of the receive array orientation, the $4 \times 2$ channel matrix ${\bf H}$ contains at least one $2 \times 2$ submatrix ${\bf H}_{\rm sub}$ such that the correlation $\mu_{\rm sub}$ between the two columns of ${\bf H}_{\rm sub}$ is at the most 
$\cos\left( {\pi}/{2\sqrt{2} \eta} \right)$. 
This latter problem is equivalent to finding the maximum distortion when a unit vector in $\mathbb{R}^3$ is quantized using a codebook $\mathcal{G}$ consisting of $12$ unit vectors that correspond to the $6$ edges of the tetrahedron along with the polarities \mbox{$\pm 1$}. The computation of this maximum distortion is then simplified by showing that $\mathcal{G}$ is a \emph{group code}~\cite{Sle_Bell_68}.

%%%%%

\begin{figure}[!t] 
\centering
\includegraphics[width=2.5in]{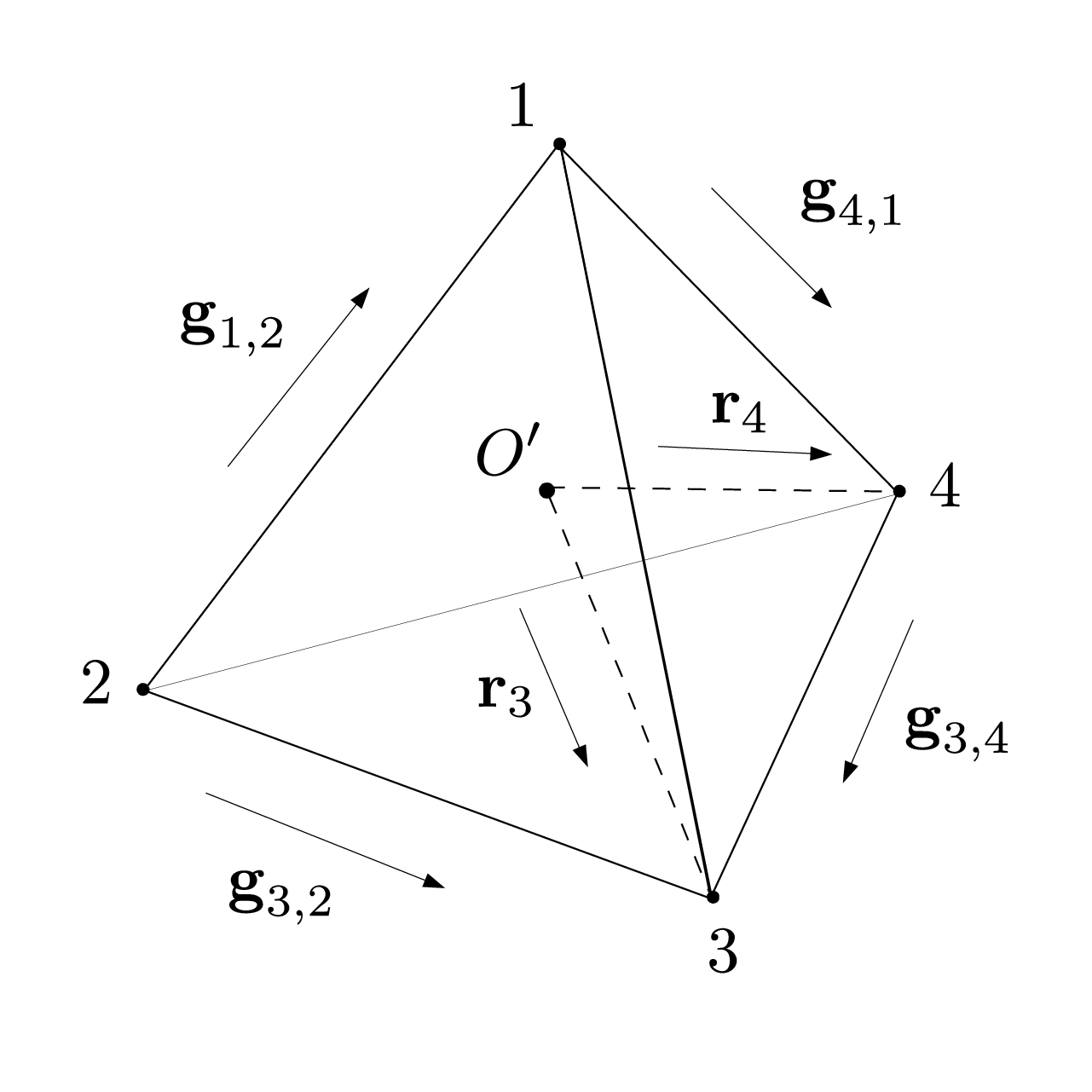}
\caption{The tetrahedron arrangement illustrating the vertices $1,\dots,4$, the reference $O'$ at the centroid of the tetrahedron, and the directions of a few of the unit vectors ${\bf r}_m$ and ${\bf g}_{m,\ell}$.}
\label{fig:tetrahedron}
\end{figure}

We first introduce some notation to capture the geometrical properties of the tetrahedral array.
Consider the tetrahedron shown in Fig.~\ref{fig:tetrahedron} with the centroid $O'$. 
Let ${\bf r}_m \in \mathbb{R}^3$ be the unit vector in the direction of the $m^{th}$ receive antenna with respect to the reference $O'$. Hence, the position vector of the $m^{th}$ receive antenna is $d_m {\bf r}_m$. 
If one applies a $3$-dimensional rotation ${\bf U} \in \mathbb{R}^{3 \times 3}$ on the receive system about $O'$, the position of the $m^{th}$ receive antenna is $d_m{\bf Ur}_m$. 
It is straightforward to show that the polar angle $\theta_m$ of the $m^{th}$ rotated receive antenna (cf. Fig.~\ref{fig:geometry_los}) satisfies $\cos\theta_m = {\bf r}_m^\intercal{\bf U}^\intercal {\bf \tilde{v}}$, where the unit vector ${\bf \tilde{v}} = \begin{bmatrix} -\sin\beta , \, 0 , \, \cos\beta \end{bmatrix}^\intercal$.
Since ${\bf U}$ is an arbitrary rotation matrix, the set of all possible values assumed by the vector ${\bf v}= {\bf U}^\intercal{\bf \tilde{v}}$ is the sphere $\mathbb{S}^2$ consisting of all unit vectors in $\mathbb{R}^3$. From~\eqref{eq:mu_model}, the correlation $\mu$ for a tetrahedral receiver is
\begin{align*} 
\mu &= \frac{1}{4} \left\vert \sum_{m=1}^{4}\exp \left( \frac{i2\pi d_t d_m \cos\beta\cos\theta_m}{R\lambda} \right) \right\vert,
\end{align*} 
where $\cos\theta_m = {\bf r}_m^\intercal{\bf U}^\intercal\tilde{\bf v} = {\bf r}_m^\intercal{\bf v}$, and ${\bf v} \in \mathbb{S}^2$ captures the effect of the rotation undergone by the receive array.
For any $m \neq \ell$, the unit vectors ${\bf r}_m$ and ${\bf r}_{\ell}$ satisfy $\| {\bf r}_m - {\bf r}_{\ell} \|=\sqrt{{8}/{3}}$. Let 
\begin{align*}
{\bf g}_{m,\ell} = \frac{{\bf r}_m - {\bf r}_\ell}{\| {\bf r}_m - {\bf r}_\ell \|} = \sqrt{\frac{3}{8}}\left( {\bf r}_m - {\bf r}_\ell \right)
\end{align*}
be the unit vector along ${\bf r}_m - {\bf r}_\ell$, i.e., along the edge of the tetrahedron between the vertices $m$ and $\ell$ (see Fig.~\ref{fig:tetrahedron}).

Let $\Hsub$ be the $2 \times 2$ submatrix of ${\bf H}$ formed using the $m^{th}$ and $\ell^{th}$ rows. Note that $\Hsub$ is the channel response seen through the receive antennas $m$ and $\ell$. Using the fact that $d_m=d_{\ell}=\sqrt{{3}/{8}} \, d_r$, the correlation between the columns of $\Hsub$ can be written as
\begin{align}
\musub &= \frac{1}{2} \Big\vert \exp \left( {\frac{i2\pi d_t d_m \cos\beta \, {\bf r}_m^\intercal{\bf v}}{R\lambda} }\right) + \nonumber \\
&~~~~~~~~~~~~~~~~~~~~~~~~~~~~~~~ \exp\left( {\frac{i2\pi d_t d_{\ell} \cos\beta \, {\bf r}_\ell^\intercal{\bf v}}{R\lambda} } \right) \Big\vert \nonumber \\
& = \frac{1}{2} \left\vert 1 + \exp\left( { \frac{i2\pi d_t d_m \cos\beta ({\bf r}_m-{\bf r}_\ell)^\intercal{\bf v}}{R\lambda} } \right) \right\vert \nonumber \\
& = \frac{1}{2} \left\vert 1 + \exp\left( { \frac{i2\pi d_t d_m \sqrt{{8}/{3}} \, \cos\beta \, {\bf g}_{m,\ell}^\intercal{\bf v}}{R\lambda} } \right) \right\vert \nonumber \\
& = \frac{1}{2} \left\vert 1 + \exp\left( { i \frac{\pi}{\eta} \, {\bf g}_{m,\ell}^\intercal{\bf v} } \right) \right\vert \nonumber \\ 
& = \left\vert \cos\left( \frac{\pi}{2\eta} \, {\bf g}_{m,\ell}^\intercal{\bf v} \right) \right\vert, \label{eq:musub_g_v}
\end{align}
where the fourth equality follows from~\eqref{eq:eta} and the last equality uses straightforward algebraic manipulations. Given an `orientation' ${\bf v}$, we intend to find the submatrix $\Hsub$ with the least correlation $\musub$. 
If \mbox{$\eta \geq 1$}, we have 
\begin{align*}
\left\vert \frac{\pi}{2\eta} \, {\bf g}_{m,\ell}^\intercal{\bf v} \right\vert \leq \frac{\pi}{2}.
\end{align*} 
Since $\cos$ is decreasing function in the interval \mbox{$[0,{\pi}/{2}]$}, from~\eqref{eq:musub_g_v}, the problem of finding $\musub$ translates to finding the edge ${\bf g}_{m,\ell}$ of the tetrahedron that has the largest inner product with ${\bf v}$. 

We will now show that for any ${\bf v} \in \mathbb{S}^2$ there exists a ${\bf g}_{m,\ell}$ such that $\sqrt{{1}/{2}} \leq {\bf g}_{m,\ell}^\intercal{\bf v} \leq 1$. Since
\begin{align*} %\label{eq:dist_inner_prod}
\| {\bf v} - {\bf g}_{m,\ell} \|^2 = \|{\bf v}\|^2 + \|{\bf g}_{m,\ell}\|^2 - 2\,{\bf g}_{m,\ell}^\intercal{\bf v} = 2 \! \left( 1 -  {\bf g}_{m,\ell}^\intercal{\bf v} \right)
\end{align*}
this is equivalent to finding the maximum squared Euclidean error when the set of vectors 
% \begin{align*}
$\mathcal{G} = \left\{ {\bf g}_{m,\ell}~|~m \neq \ell \right\}$
% \end{align*}
is used as a codebook for quantizing an arbitrary unit vector ${\bf v}$ in $\mathbb{R}^3$. The set $\mathcal{G}$ contains $12$ vectors, corresponding to the $6$ edges of the tetrahedron together with the polarity $\pm 1$. 

\begin{proposition} \label{prop:geometry}
For any ${\bf v} \in \mathbb{S}^2$, there exist $m,\ell \in \{1,2,3,4\}$, $m \neq \ell$, such that ${\bf g}_{m,\ell}^\intercal{\bf v} \geq \sqrt{{1}/{2}}$.
\end{proposition}
\begin{IEEEproof}
With some abuse of notation we will denote the elements of $\mathcal{G}$ as ${\bf g}_1,\dots,{\bf g}_{12}$. For each $i=1,\dots,12$, let 
\begin{align} \label{eq:inner_prod_regions}
\mathcal{D}_i = \left\{ {\bf v} \in \mathbb{S}^2  \,| \, {\bf g}_i^\intercal {\bf v} \geq {\bf g}_j^\intercal {\bf v}, \textrm{ for all } j \neq i \right\}
\end{align}
be the set of unit vectors that are closer to ${\bf g}_i$ than any other ${\bf g}_j \in \mathcal{G}$. Since $\cup_{i} \mathcal{D}_i = \mathbb{S}^2$, it is enough to show that
\begin{align*}
\min_{i} \min_{{\bf v} \in \mathcal{D}_i} {\bf g}_i^\intercal{\bf v} = \sqrt{\frac{1}{2}}.
\end{align*}

As we now show, the regions $\mathcal{D}_1,\dots,\mathcal{D}_{12}$ are congruent to each other. Let $\mathcal{H}$ be the symmetry group of the tetrahedron, i.e., the set of all orthogonal transformations on $\mathbb{R}^3$ that map the tetrahedron onto itself. It is known that the group $\mathcal{H}$ is isomorphic to the symmetric group $\mathcal{S}_4$ of degree $4$, and every element of $\mathcal{H}$ is uniquely identified by its action on the set of vertices, which is isomorphic to the action of the corresponding element in $\mathcal{S}_4$ on the set $\{1,2,3,4\}$; see~\cite{Mil_Acad_73}. Since for any two given pairs $(m_1,\ell_1)$ and $(m_2,\ell_2)$, with $m_1 \neq \ell_1$ and $m_2 \neq \ell_2$, there exists a permutation on $\{1,2,3,4\}$ that maps $m_1$ to $m_2$ and $\ell_1$ to $\ell_2$, we see that there exists an orthogonal transformation ${\bf M} \in \mathcal{H}$ such that
\begin{align*}
{\bf r}_{m_2} = {\bf Mr}_{m_1} \textrm{ and } {\bf r}_{\ell_2} = {\bf Mr}_{\ell_1}.
\end{align*}
This can be extended to a group action on $\mathcal{G}$ as 
\begin{align*}
{\bf Mg}_{m_1,\ell_1} &= {\bf M}\left( \frac{{\bf r}_{m_1} - {\bf r}_{\ell_1}}{\|{\bf r}_{m_1} - {\bf r}_{\ell_1}\|} \right) 
                      = \sqrt{\frac{3}{8}} {\bf M} \left( {\bf r}_{m_1} - {\bf r}_{\ell_1}\right) \\
                      &= \sqrt{\frac{3}{8}} \left( {\bf r}_{m_2} - {\bf r}_{\ell_2}\right)
                      = {\bf g}_{m_2,\ell_2}.
\end{align*}
Thus we see that the group $\mathcal{H}$ acts transitively on $\mathcal{G}$, i.e.,
\begin{align*}
\mathcal{G} = \left\{ {\bf Mg}_i \, \vert \, {\bf M} \in \mathcal{H} \right\} \textrm{ for every } i=1,\dots,12.
\end{align*}
This makes $\mathcal{G}$ a group code, and consequently, the regions $\mathcal{D}_1,\dots,\mathcal{D}_{12}$ are congruent to each other~\cite{Sle_Bell_68}, i.e., for every \mbox{$1 \leq i < j \leq 12$}, there exists an orthogonal transformation ${\bf M} \in \mathcal{H}$ such that
\begin{align*}  
\mathcal{D}_j = {\bf M} \mathcal{D}_i = \left\{ {\bf Mv} \, \vert \, {\bf v} \in \mathcal{D}_i  \right\}.
\end{align*}
Since orthogonal transformations conserve inner products and since ${\bf g}_i \in \mathcal{D}_i$ for all $i$, we have
\begin{align*}
\min_{{\bf v} \in \mathcal{D}_i} {\bf g}_i^\intercal{\bf v} = \min_{{\bf v} \in \mathcal{D}_j} {\bf g}_j^\intercal{\bf v} \textrm{ for any } i \neq j. 
\end{align*}
Thus, to complete the proof it is enough to show that
\begin{align*}
\min_{{\bf v} \in \mathcal{D}_1} {\bf g}_1^\intercal{\bf v} = \sqrt{\frac{1}{2}}.
\end{align*}

\begin{figure}[!t] 
\centering
\includegraphics[width=3.4in]{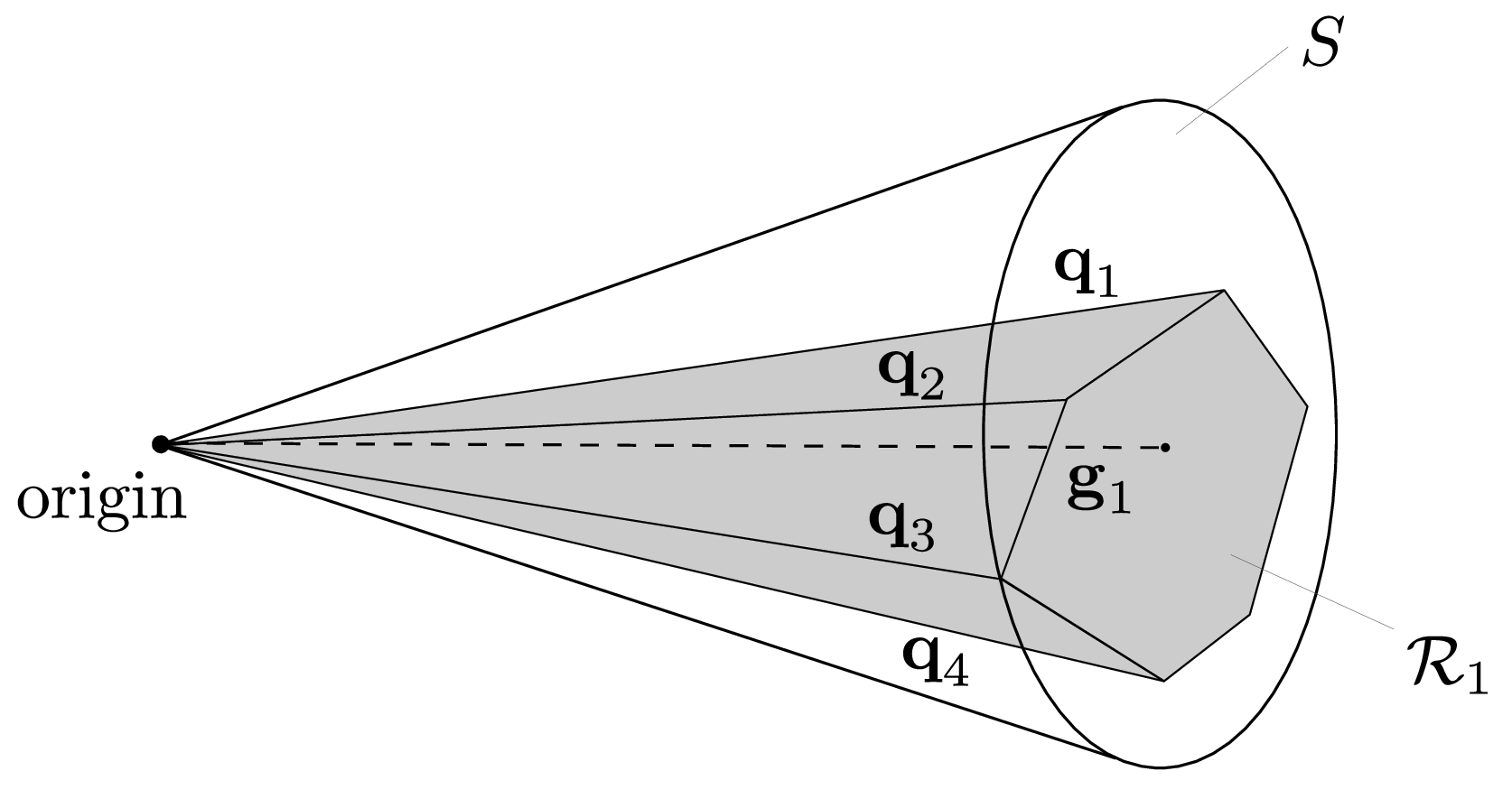}
\caption{An illustration of the cones $S$ and $\mathcal{R}_1$ used in the proof of Proposition~\ref{prop:geometry}. The cone $S$ is circular with axis ${\bf g}_1$ (dashed line). The cone $\mathcal{R}_1$ is bounded by hyperplanes, and its edges are along the vectors ${\bf q}_1,\dots,{\bf q}_6$. The edge ${\bf q}_3$ is the farthest from the axis ${\bf g}_1$ and lies on the surface of $S$.}
\label{fig:illustration_of_cones}
\end{figure}

We now restrict ourselves to one particular region $\mathcal{D}_1$ and find the smallest value of ${\bf g}_1^\intercal{\bf v}$. Note that when ${\bf v} \in \mathbb{S}^2$, the inner product of ${\bf v}$ with ${\bf g}_i$ decreases with increasing distance $\| {\bf v} - {\bf g}_i \|$. Thus, from~\eqref{eq:inner_prod_regions}, $\mathcal{D}_1$ is the intersection of $\mathbb{S}^2$ with the set of all points in $\mathbb{R}^3$ that are closer to ${\bf g}_1$ than any other ${\bf g}_i \in \mathcal{G}$. The region $\mathcal{D}_1$ is called a \emph{fundamental region} of the group code $\mathcal{G}$ and is bounded by two-dimensional planes passing through the origin~\cite{Sle_Bell_68}. The half-spaces $\mathcal{P}_i$ that define this fundamental region are 
\begin{align*}
\mathcal{P}_i &= \left\{ {\bf x} \in \mathbb{R}^3 \, \vert \, \| {\bf x} - {\bf g}_1 \| \leq \| {\bf x} - {\bf g}_i \|  \right\} \\ 
              &= \left\{ {\bf x} \in \mathbb{R}^3 \, \vert \, {\left( {\bf g}_1 - {\bf g}_i \right)}^\intercal {\bf x} \geq 0  \right\},
\end{align*}
and are related to $\mathcal{D}_1$ as
\begin{align*}
\mathcal{D}_1 &= \mathbb{S}^2 \cap \mathcal{R}_1, \textrm{ where } \mathcal{R}_1 = \cap_{i=2}^{12} \mathcal{P}_i.
\end{align*}
The group code $\mathcal{G}$ and the $11$ half-spaces $\mathcal{P}_i$ can be explicitly calculated starting from the geometry of the tetrahedron, and it can be verified that $\mathcal{R}_1$, and hence $\mathcal{D}_1$, is bounded by exactly $6$ planes arising from $6$ of the eleven half-spaces $\mathcal{P}_i$. 
The region $\mathcal{R}_1$ is a convex cone~\cite{Sle_Bell_68} generated from the $6$ edges running along the vectors ${\bf q}_1,\dots,{\bf q}_6$ that are the intersections between the $6$ hyperplanes, i.e., $\mathcal{R}_1$ is the infinite cone generated from the convex hull of the set $\{ {\bf q}_1,\dots,{\bf q}_6 \}$. Fig.~\ref{fig:illustration_of_cones} shows an illustration of the geometry considered in this proof (the depiction of ${\bf q}_1,\dots,{\bf q}_6$ is not exact).
Since 
\begin{align} \label{eq:angle_cone}
\min_{{\bf v} \in \mathcal{D}_1} {\bf g}_1^\intercal {\bf v} = \min_{{\bf x} \in \mathcal{R}_1} \frac{{\bf g}_1^\intercal {\bf x}}{\|{\bf x}\|},
\end{align}
and since ${{\bf g}_1^\intercal {\bf x}}/{\|{\bf x}\|}$ is the cosine of the angle between ${\bf x}$ and ${\bf g}_1$, our problem is to find a vector in $\mathcal{R}_1$ which makes the largest angle with ${\bf g}_1$. The set of points that make a constant angle with ${\bf g}_1$ form the surface of an infinite circular cone with ${\bf g}_1$ as its axis. Thus~\eqref{eq:angle_cone} is equivalent to finding the smallest circular cone $S$, with ${\bf g}_1$ as the axis, that contains the conical region $\mathcal{R}_1$. Since $\mathcal{R}_1$ is generated by ${\bf q}_1,\dots,{\bf q}_6$, $S$ is the smallest circular cone that contains the vectors ${\bf q}_1,\dots,{\bf q}_6$, and has ${\bf g}_1$ as the axis. It follows that $S$ contains on its surface the vector ${\bf q}_i$, from among ${\bf q}_1,\dots,{\bf q}_6$, that makes the largest angle with ${\bf g}_1$. Thus,
\begin{align*}
\min_{{\bf v} \in \mathcal{D}_1} {\bf g}_1^\intercal {\bf v} &= \min_{{\bf x} \in \mathcal{R}_1} \frac{{\bf g}_1^\intercal {\bf x}}{\|{\bf x}\|} 
                                                             = \min_{{\bf x} \in {S}} \frac{{\bf g}_1^\intercal {\bf x}}{\|{\bf x}\|} 
\end{align*}
The numerical value $\min_{i \in \{1,\dots,6\}} \, {{\bf g}_1^\intercal {\bf q}_i}\,/\,{\|{\bf q}_i\|} = {1}/{\sqrt{2}}$ is obtained by a direct computation of the half-spaces $\mathcal{P}_1,\dots,\mathcal{P}_{11}$, and the resulting vectors ${\bf q}_1,\dots,{\bf q}_6$ arising from the tetrahedral geometry.
\end{IEEEproof}

\begin{proposition} \label{prop:musub_tetr}
If a tetrahedral array is used at the receiver and \mbox{$\eta \geq 1$}, then for every receive orientation ${\bf U}$, there exists a \mbox{$2 \times 2$} submatrix $\Hsub$ of the channel matrix ${\bf H}$ such that
\begin{align*}
0 \leq \musub \leq \cos\left( \frac{\pi}{2\sqrt{2} \eta} \right),
\end{align*}
where $\musub$ is the correlation between the two columns of $\Hsub$.
\end{proposition}
\begin{IEEEproof}
From Proposition~\ref{prop:geometry}, there exist $m \neq \ell$ such that ${\bf g}_{m,\ell}^\intercal{\bf v} \geq \sqrt{{1}/{2}}$. Let ${\bf H}_{\rm sub}$ be the submatrix of ${\bf H}$ formed by the $m^{th}$ and $\ell^{th}$ rows. From~\eqref{eq:musub_g_v} and the hypothesis that $\eta \geq 1$, we have
% \begin{align*}
$\musub = \left\vert \cos\left( \frac{\pi}{2\eta} \, {\bf g}_{m,\ell}^\intercal{\bf v} \right) \right\vert \leq \cos\left( \frac{\pi}{2\eta} \, \sqrt{\frac{1}{2}} \right)$.
% \end{align*} 
\end{IEEEproof}

The following upper bound on $\mu^*(\eta)$ follows immediately from Proposition~\ref{prop:musub_tetr}.

\begin{theorem} \label{thm:bounds_mu_tetr}
For a tetrahedral receive array and $\eta \geq 1$,
\begin{align*}
\mu^*(\eta) \leq \frac{1}{2}\left( 1 + \cos\left( \frac{\pi}{2\sqrt{2} \eta} \right) \right).
\end{align*}
\end{theorem}
\begin{IEEEproof}
Let \mbox{${\bf H}=[h_{m,n}]$} be the $4 \times 2$ channel matrix. From Proposition~\ref{prop:musub_tetr}, assume without loss of generality that the \mbox{$2 \times 2$} submatrix formed from the first two rows has correlation $\musub \leq \cos\left( {\pi}/{2\sqrt{2}\eta} \right)$. Then,
\begin{align*}
 \mu &= \frac{1}{4} \left\vert h_{1,1}^\dag h_{1,2} + h_{2,1}^\dag h_{2,2} + h_{3,1}^\dag h_{3,2} + h_{4,1}^\dag h_{4,2} \right\vert \\
     &\leq \frac{1}{4} \left\vert h_{1,1}^\dag h_{1,2} + h_{2,1}^\dag h_{2,2} \right\vert + \frac{1}{4} \left\vert h_{3,1}^\dag h_{3,2} + h_{4,1}^\dag h_{4,2} \right\vert \\
     &= \frac{1}{2} \musub + \frac{1}{4} \left\vert h_{3,1}^\dag h_{3,2} + h_{4,1}^\dag h_{4,2} \right\vert \\
     &\leq \frac{1}{2} \cos\left( \frac{\pi}{2\sqrt{2} \eta} \right) + \frac{2}{4},
\end{align*} 
where the last inequality follows from Proposition~\ref{prop:musub_tetr} and the fact that all $h_{m,n}$ have unit magnitude.
\end{IEEEproof}

The upper bound $\left( 1 + \cos\left( {\pi}/{2\sqrt{2}} \right) \right)/2$ on $\mu^*(\eta)$ is less than $1$ for \mbox{$\eta \geq 1$}. Since ${\sf d}(\mu,\Delta{\bf X})$ is a decreasing function of $\mu$, we have \mbox{${\sf d}(\mu^*(\eta),\Delta{\bf X}) > {\sf d}(1,\Delta{\bf X})$}. Hence, the geometry of the tetrahedral arrangement allows the error probability to decay faster than that of rank $1$ LoS MIMO channels, and provides performance improvement over any planar arrangement $n_r=4$ of antennas, irrespective of the code used at the transmitter. 
Note that this gain of the tetrahedral arrangement over planar arrays is not due to larger inter-antenna distances $d_t$ and $d_r$.

From~\eqref{eq:cauchy_schwarz}, we have \mbox{${\sf d}(1,\Delta{\bf X}) \geq \left( \|\Delta{\bf x}_1\| - \|\Delta{\bf x}_2\|\right)^2$}. Using $\mu^*<1$, we obtain
\begin{align*}
{\sf d}(\mu^*,\Delta{\bf X}) > {\sf d}(1,\Delta{\bf X}) \geq \left( \|\Delta{\bf x}_1\| - \|\Delta{\bf x}_2\|\right)^2 \geq 0.
\end{align*} 
Hence, unlike the planar case, the error probability of a tetrahedral receiver is exponential in ${\sf SNR}$ for any code $\mathscr{C}$. 

\begin{example} \label{ex:vblast:2}
\emph{Performance of Spatial Multiplexing with Tetrahedral Receive Array}.
Consider the SM scheme signalled over $n_t=2$ antennas using $4$-QAM symbols. Let the transmit orientation $\beta=0$ be fixed, the inter-terminal distance $R=10$~m, $\lambda=4.2$~mm, and $d_t=d_r=0.145$~m. Then, $\eta={R\lambda}/({2d_td_r \, \cos\beta})=1$, and from Theorem~\ref{thm:bounds_mu_tetr}, $\mu^*(\eta) \leq 0.722$. An exhaustive numerical computation over all pairs of codewords yields $\min_{\Delta{\bf X}}{\sf d}(0.722,\Delta{\bf X})=0.556$. Using~\eqref{eq:PEP_eta}, the pairwise error probability of SM for fixed transmit orientation and random receive orientation can be upper bounded as 
\begin{align*}
\mathbb{E}({\sf PEP}) \leq {\sf PEP}^* &\leq \frac{1}{2} \exp\left( -{\sf SNR} \, \mu^*(1) \right) \\
                                       &\leq \frac{1}{2}\exp\left( -{\sf SNR} \times 0.556 \right).
\end{align*} 
On the other hand, as shown in Example~\ref{ex:vblast}, for any planar receiver array, the error rate is not better than ${\sf SNR}^{-3}$.
\end{example}

\subsection{System design for arbitrary array orientations} \label{sec:system_design}

In Section~\ref{sec:3A}, we assumed that $\eta$ was fixed, i.e., the transmit orientation $\beta$ and inter-terminal distance $R$ were fixed, and we studied the effect of an arbitrary rotation ${\bf U}$ of the receive array on $\mu$ and error probability. We now design a system that allows arbitrary transmit and receive array orientations and a range of values $R_{\min} \leq R \leq R_{\max}$. 
It is desirable that the LoS MIMO system guarantees a minimum channel quality i.e., \mbox{$\mu \leq \mu_{\max}$}, for some \mbox{$\mu_{\max}<1$}.
Using~\eqref{eq:PEP_eta}, for such a system,
\begin{align*}
\mathbb{E}({\sf PEP}) \leq {\sf PEP}^* \leq \frac{1}{2} \exp\left( -\frac{n_r\,{\sf SNR}}{4} \, {\sf d}(\mu_{\max},\Delta{\bf X}) \right).
\end{align*} 
Using union bound, the average codeword error rate and bit error rate of the system can be upper bounded by
\begin{align*}
\frac{|\mathscr{C}|}{2} \exp\left(-\frac{n_r\,{\sf SNR}}{4} \, \min_{\Delta{\bf X}} {\sf d}(\mu_{\max},\Delta{\bf X})  \right).
\end{align*} 
Hence, the coding gain of an arbitrary coding scheme $\mathscr{C}$ over this LoS MIMO system is $\min_{\Delta{\bf X}}{\sf d}(\mu_{\max},\Delta{\bf X})$.

When the number of transmit antennas \mbox{$n_t=2$}, by choosing \mbox{$\beta={\pi}/{2}$}, we observe from~\eqref{eq:mu_model} that the worst case correlation \mbox{$\mu_{\max}=1$} irrespective of the array geometry used at the receiver.
Hence, in order to have \mbox{$\mu_{\max}<1$}, we need more than $2$ antennas at the transmitter. 

Suppose the transmitter uses an array of \mbox{$n_t \geq 3$} antennas. Based on the transmit array orientation, one can choose $2$ of the $n_t$ antennas for signal transmission so that the angle $\beta$ corresponding to the chosen pair of antennas is minimum. 
\begin{figure}[!t]
\centering
\includegraphics[width=1.5in]{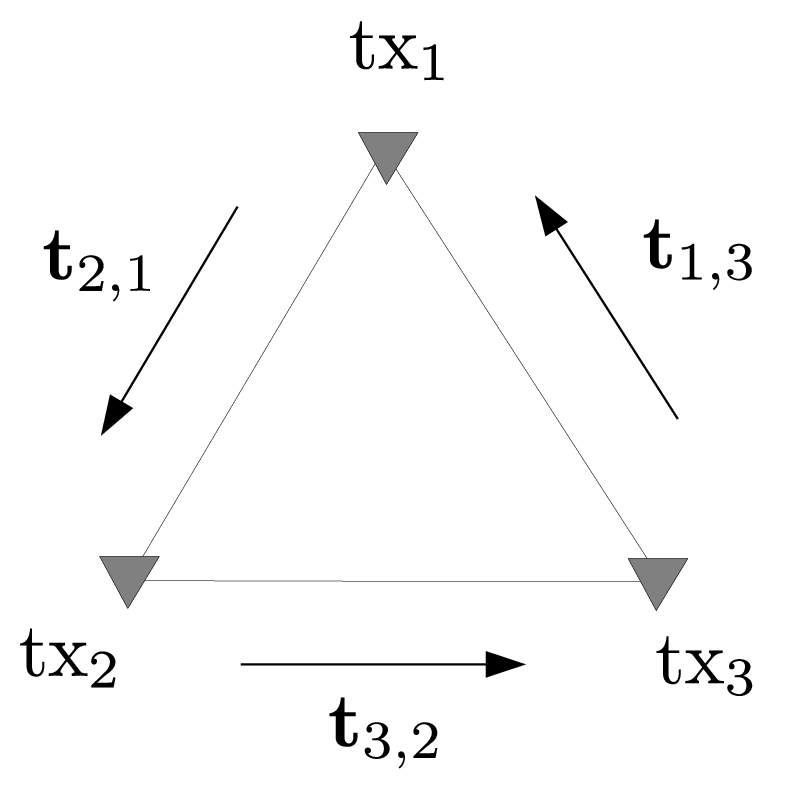}
\caption{Triangular arrangement of transmit antennas.}
\label{fig:transmit_traingle}
\end{figure} 
For example, let \mbox{$n_t=3$} antennas be placed at the vertices of an equilateral triangle with inter-antenna distance $d_t$, as shown in Fig.~\ref{fig:transmit_traingle}.
Let ${\bf t}_{m,n}$ be the unit vector in $\mathbb{R}^3$ in the direction of the position of transmit antenna $m$ with respect to the position of transmit antenna $n$. 
Note that the vectors ${\bf t}_{m,n}$ vary with changes in the transmit array orientation.
If antennas $m$ and $n$ are used for transmission and if ${\bf u} \in \mathbb{R}^3$ is the unit vector along the direction $OO'$ of transmission, then \mbox{$\sin\beta = {\bf u}^\intercal{\bf t}_{m,n}$} (cf. Fig.~\ref{fig:geometry_los}, where ${\rm tx}_1$ and ${\rm tx}_2$ correspond to ${\rm tx}_m$ and ${\rm tx}_n$, respectively). The six vectors in the set 
\begin{align*}
\mathcal{T}=\{{\bf t}_{m,n} \, \vert \, m,n=1,2,3, \, m \neq n\}
\end{align*} 
are arranged symmetrically in a two-dimensional plane at regular angular intervals of ${\pi}/{3}$. Let ${\bf u}_{\parallel}$ and ${\bf u}_{\perp}$ be the components of ${\bf u}$ parallel and perpendicular to the plane of $\mathcal{T}$, respectively. Since the vectors in $\mathcal{T}$ divide the plane into regular conical regions of angular width ${\pi}/{3}$, there exists at least one vector ${\bf t}_{m,n} \in \mathcal{T}$ such that the angle between ${\bf t}_{m,n}$ and ${\bf u}_{\parallel}$ lies in the interval $[- {\pi}/{6}, + {\pi}/{6}]$, i.e., 
\begin{align*}
\frac{\vert {\bf u}_{\parallel}^\intercal{\bf t}_{m,n} \vert}{\|{\bf u}_{\parallel}\|} \leq \sin\left( \frac{\pi}{6} \right) = \frac{1}{2}.
\end{align*}
We can thus upper bound $|{\bf u}^\intercal{\bf t}_{m,n}|^2$ as follows
\begin{align*}
|{\bf u}^\intercal{\bf t}_{m,n}|^2 &= |{\bf u}_{\perp}^\intercal{\bf t}_{m,n}|^2 + |{\bf u}_{\parallel}^\intercal{\bf t}_{m,n}|^2  \leq 0 + \|{\bf u}_{\parallel}\|^2 \frac{1}{4} \leq \frac{1}{4}.
\end{align*}
Thus there exists a ${\bf t}_{m,n}$ such that 
%% \begin{align*}
$\displaystyle |\sin(\beta)| = |{\bf u}^\intercal{\bf t}_{m,n}| \leq {1}/{2}$, 
%% \end{align*}
i.e., \mbox{$\beta \in [-{\pi}/{6},{\pi}/{6}]$}.
Hence, if the transmit array is an equilateral triangle, by appropriately choosing $2$ out of the $3$ available antennas for signalling, one can ensure $\vert\beta\vert \leq {\pi}/{6}$. 

\begin{figure}[!t]
\centering
\includegraphics[width=3.4in]{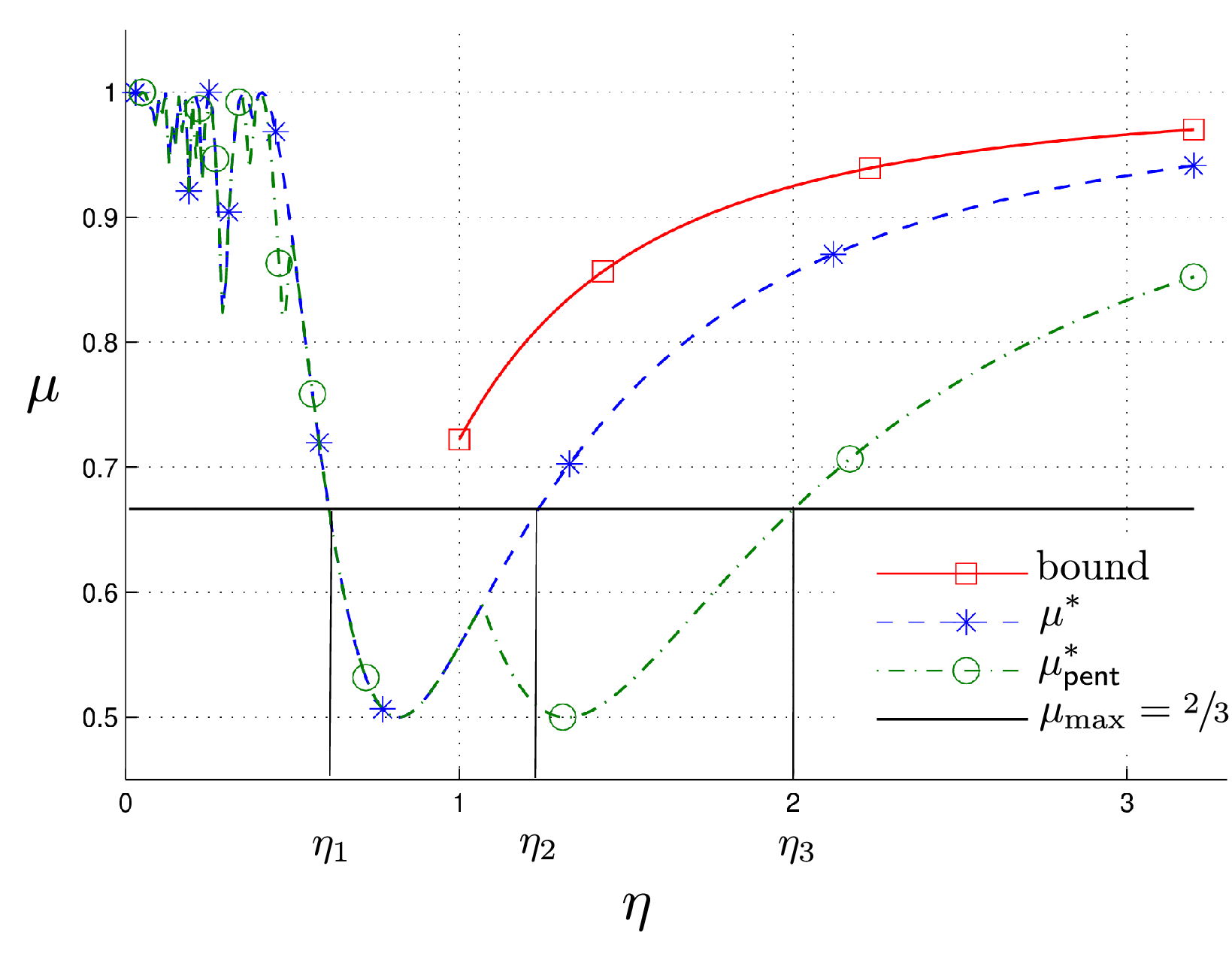}
\caption{The functions $\mu^*$, $\mu_{\sf pent}^*$, upper bound on $\mu^*$ and the line $\mu_{\rm max}={2}/{3}$.}
\label{fig:mu_pentagon}
\end{figure}

The upper bound on $\mu^*(\eta)$ of Theorem~\ref{thm:bounds_mu_tetr} is not tight and is available only for \mbox{$\eta \geq 1$}. Since this bound can not be used as a good estimate of $\mu^*(\eta)$ and the analytical computation of the exact expression~\eqref{eq:mu_star} of $\mu^*(\eta)$ appears to be difficult, we use numerically computed values of $\mu^*(\eta)$ for system design. The function $\mu^*(\eta)$ and the upper bound of Theorem~\ref{thm:bounds_mu_tetr} are shown in Fig.~\ref{fig:mu_pentagon}.
Using the exact function $\mu^*(\eta)$, the requirement on channel quality \mbox{$\mu \leq \mu_{\max}$} can be translated into a criterion \mbox{$\eta \in [\eta_{\min},\eta_{\max}]$}. From~\eqref{eq:eta}, for fixed $d_t$, $d_r$, $\lambda$, and $\vert\beta\vert \leq \beta_{\max}$, we have
\begin{equation} \label{eq:eta_max_min}
\eta_{\min}= \frac{R_{\min} \lambda}{2 d_t d_r} \textrm{ and } \eta_{\max} = \frac{R_{\max} \lambda}{2 d_t d_r \cos{\beta_{\max}}}.
\end{equation}
The range $[R_{\min},R_{\max}]$ can thus be obtained from~\eqref{eq:eta_max_min}.

\begin{example} \label{ex:triangle}
Suppose we require $\mu_{\max}= {2}/{3}$ with $\lambda=4.2$~mm. Using a triangular transmit array we have $\beta_{\max} = {\pi}/{6}$. From Fig.~\ref{fig:mu_pentagon}, the criterion \mbox{$\mu^*(\eta) \leq {2}/{3}$} is equivalent to $\eta_{\min}=\eta_1=0.62$ and $\eta_{\max}=\eta_2=1.22$.
If each side of the triangular transmit array has length \mbox{$d_t=6$~cm}, and the tetrahedral receive array has \mbox{$d_r=25$~cm}, then from~\eqref{eq:eta_max_min} we have $R_{\min}=4.43$~m and $R_{\max}=7.75$~m.
\end{example}

\begin{figure}[!t]
\centering
\includegraphics[totalheight=1.75in,width=3.4in]{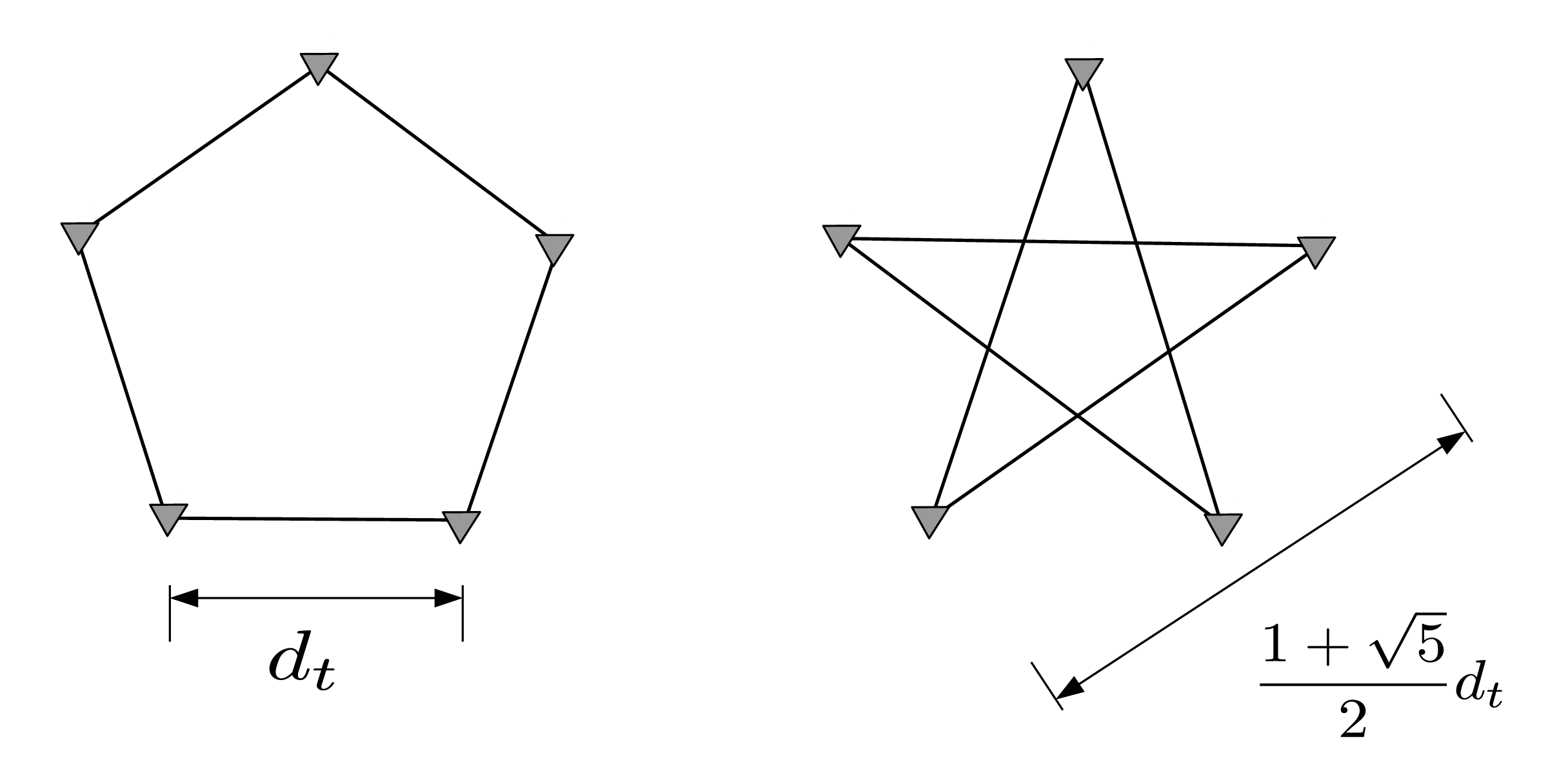}
\caption{Left: any pair of neighbouring antennas in a pentagonal array has an inter-antenna distance of $d_t$. Right: Any pair of non-neighbouring antennas has distance ${(1+\sqrt{5})d_t}/{2}$.}
\label{fig:pentagon}
\end{figure}

The narrow range of $[R_{\min},R_{\max}]$ in Example~\ref{ex:triangle} can be attributed to the small value of \mbox{${\eta_2}-{\eta_1}$} in Fig.~\ref{fig:mu_pentagon}. 
This can be improved by using a pentagonal transmit array as follows. As shown in Fig.~\ref{fig:pentagon}, with a regular pentagon, the choice of the transmit antenna pair can be divided into the following two cases:
\begin{inparaenum}[(i)]
\item the two antennas are the neighbouring vertices of the pentagon with inter-antenna distance equal to the length $d_t$ of the edge of the regular pentagon, or
\item the antennas are non-neighbouring with inter-antenna distance ${\left(1+\sqrt{5}\right)d_t}/{2}$.
\end{inparaenum}
Irrespective of the class from which the antenna pair is chosen, it is straightforward to show that $\vert\beta\vert \leq {\pi}/{10}$ can be always guaranteed. While the value of $\eta$ for the first case is given by~\eqref{eq:eta}, in the second case it reduces by a factor of ${\left(1+\sqrt{5} \right)}/{2}$ because of the larger inter-antenna distance. Thus the maximum correlation with pentagonal transmit array is
\begin{align*}
 \mu_{\sf pent}^*(\eta) = \min \left\{ \mu^*(\eta) ,\mu^*\left(\frac{2\eta}{1+\sqrt{5}}\right) \right\},
\end{align*} 
where $\mu^*(\eta)$ is given in~\eqref{eq:mu_star}. From Fig.~\ref{fig:mu_pentagon}, the value of $\eta_{\max}$ improves from $\eta_2$ to $\eta_3$, thereby widening $[R_{\min},R_{\max}]$.

\begin{example} \label{ex:pentagon}
As in Example~\ref{ex:triangle}, let \mbox{$\mu_{\max}={2}/{3}$}, \mbox{$\lambda=4.2$~mm}, \mbox{$d_t=6$~cm} and \mbox{$d_r=25$~cm}. With a pentagonal transmit array, \mbox{$\beta_{\max}={\pi}/{10}$}, and using the function $\mu_{\sf pent}^*$, we have \mbox{$\eta_{\min}=\eta_1=0.62$} and \mbox{$\eta_{\max}=\eta_3=2$}. Using~\eqref{eq:eta_max_min}, \mbox{$R_{\min}=4.43$~m} and \mbox{$R_{\max}=12.7$~m}.
\end{example}

\section{Simulation Results} \label{sec:5}

We use the system parameters $\lambda$, $d_t$, $d_r$, $R_{\max}$ and $R_{\min}$ from Example~\ref{ex:pentagon}. We assume that the transmit and receive arrays undergo independent uniformly random $3$-dimensional rotations about their centroids, and the distance $R$ between the terminals is uniformly distributed in \mbox{$[R_{\min},R_{\max}]$}. 
In all the simulations the channel matrix  ${\bf H}$ was synthesized using~\eqref{eq:h_mn} and the exact distances $\{r_{m,n}\}$ between the transmit and the receive antennas.
We consider the following three coding schemes with the transmission rate of $4$ bits per channel use: 
\begin{inparaenum}[(\itshape i\upshape)]
\item the Golden code~\cite{BRV_JIT_05} using $4$-QAM alphabet,
\item spatial multiplexing (SM)~\cite{Fos_Bell_96,WFGV_ISSSE_98,Tel_Eur_99} with $4$-QAM, and
\item uncoded $16$-QAM transmitted using only one transmit antenna (single-input multiple-output SIMO).
\end{inparaenum}
Gray mapping is used at the transmitter to map information bits to constellation points, and unless otherwise stated, maximum-likelihood (ML) decoding is performed at the receiver.
While we used pairwise error probability for performance analysis in Sections~\ref{sec:2},~\ref{sec:3} and~\ref{sec:4}, we simulate the bit error rate to compare the average error performance.

\subsection{Error performance with $n_r=4$} \label{sec:5:ml}

\begin{figure}[!t]
\centering
\includegraphics[width=3.4in]{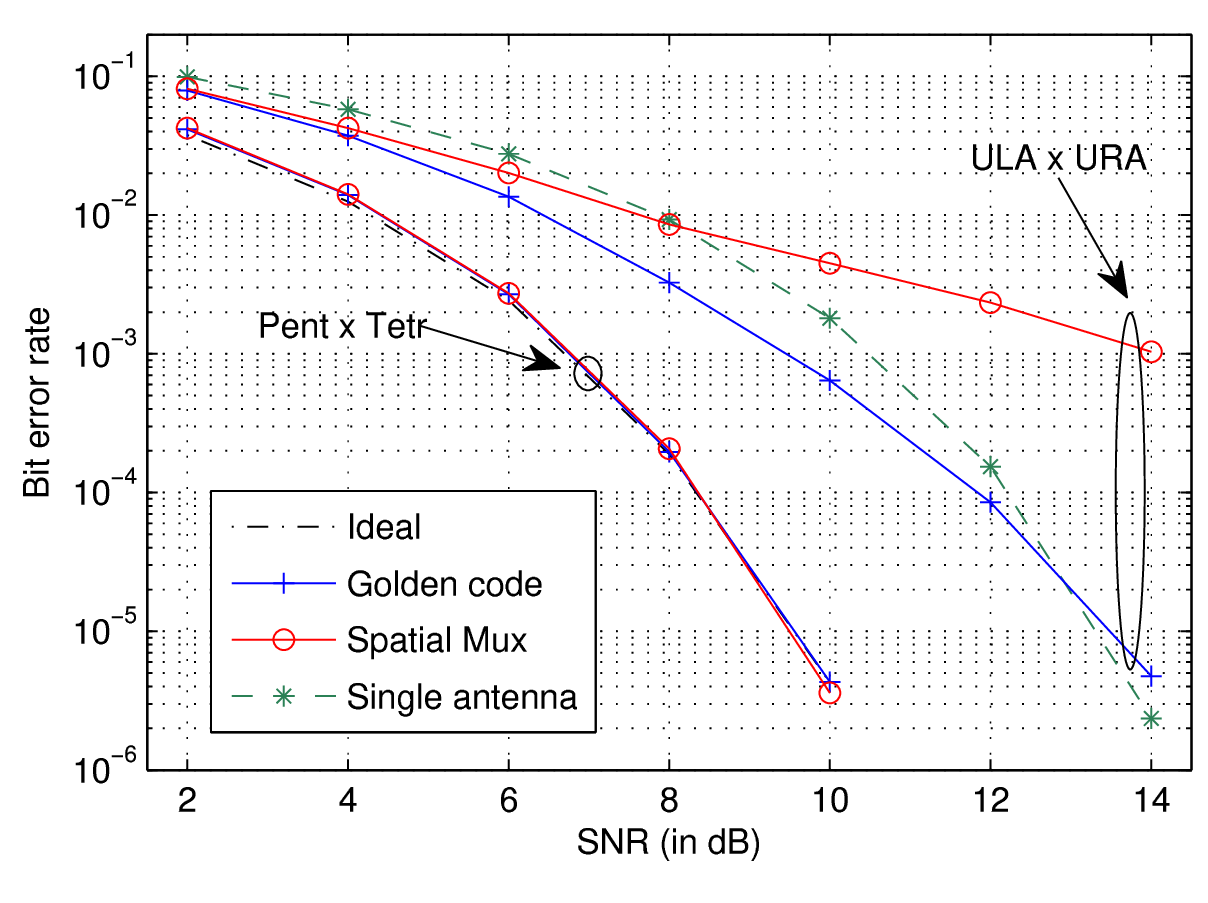}
\vspace{-3mm}
\caption{Comparison of \mbox{Pent$\times$Tetr} with \mbox{ULA$\times$URA}.}
\label{fig:manuscript1}
\end{figure} 

Fig.~\ref{fig:manuscript1} shows the performance of the three schemes with two different antenna geometries: 
\begin{inparaenum}[(\itshape i\upshape)]
\item uniform linear array (ULA) at the transmitter with \mbox{$n_t=2$}, and uniform rectangular array (URA) at receiver\footnote{The performance of uniform linear array at receiver is worse than that of URA, and hence has been omitted.} with \mbox{$n_r=4$},
\item selecting $2$ antennas from a pentagonal array at the transmitter, and using a tetrahedral array at the receiver.
\end{inparaenum}
The values of $d_t$, $d_r$ are ideal for the ULA$\times$URA configuration~\cite{BOO_Eur_07} at the distance \mbox{$R={2d_td_r}/{\lambda}=7.14$~m}, which is near the mid-point of the interval $[R_{\min},R_{\max}]$.
The performance of the single-antenna transmission scheme is independent of the receive antenna geometry since, from~\eqref{eq:h_mn}, all the channel gains of the SIMO channel have unit magnitude. %the multiple receive antennas provide only an array gain. 
Also, Fig.~\ref{fig:manuscript1} shows the performance of the ideal channel with \mbox{$\mu=0$}, i.e., \mbox{${\bf R}=\sqrt{n_r} \, {\bf I}_2$}, which is a pair of parallel AWGN channels each carrying a $4$-QAM symbol. 
From Fig.~\ref{fig:manuscript1}, we see that, with ULA$\times$URA, the performance of both SM and the Golden code are worse than SIMO at high ${\sf SNR}$. Further, since \mbox{$\min_{\Delta{\bf X}}{\sf d}(1,\Delta{\bf X})=0$} for SM, the error probability decays slowly with ${\sf SNR}$, confirming our theoretical results. With the proposed pentagon$\times$tetrahedron geometry both codes show improved performance, close to that of the ideal channel. 

\begin{figure}[!t]
\centering
\includegraphics[width=3.4in]{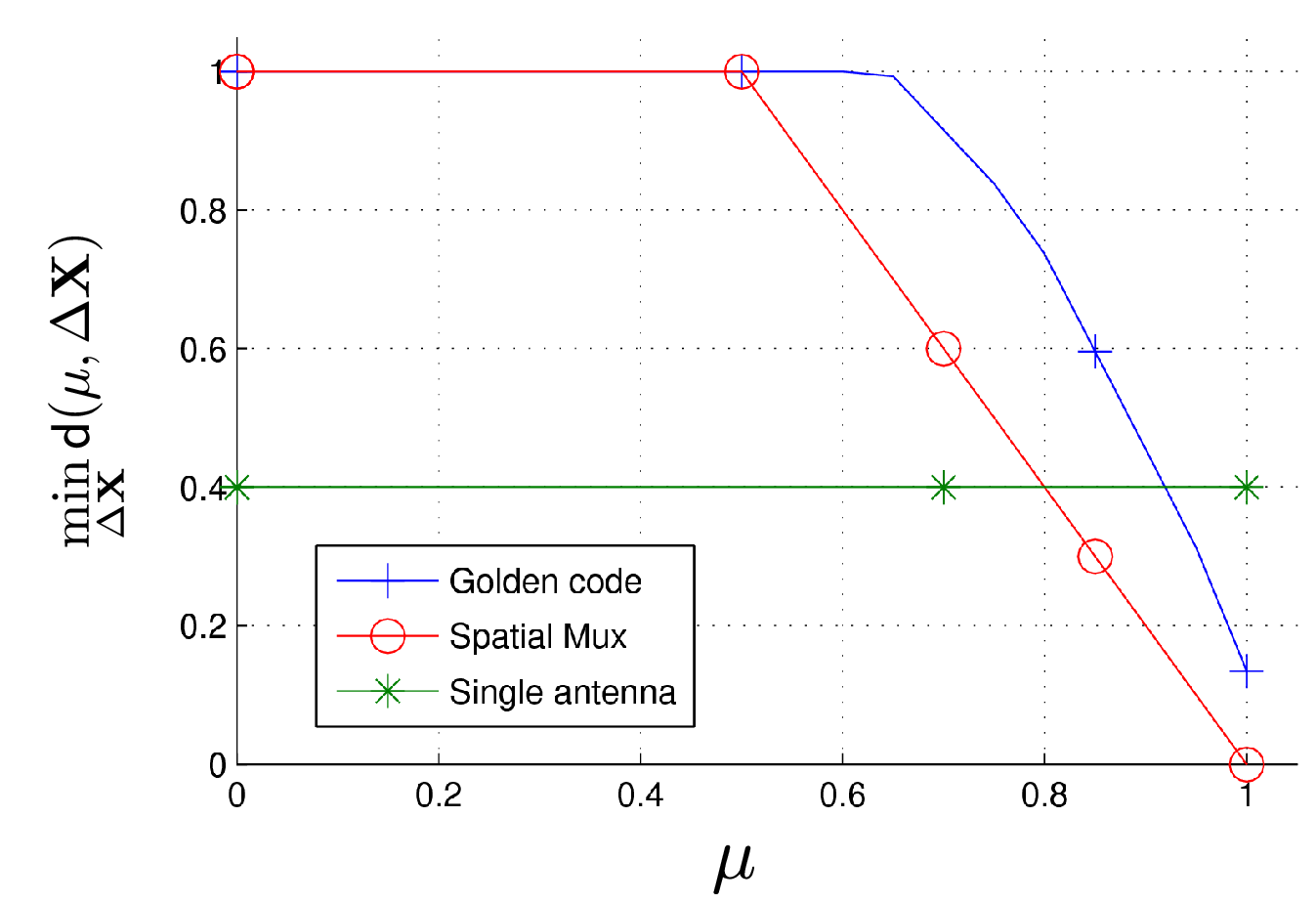}
\vspace{-3mm}
\caption{Coding gain for bit rate of $4$ bits per channel use.}
\label{fig:coding_gain}
\end{figure}

\begin{figure}[!t]
\centering
\includegraphics[width=3.4in]{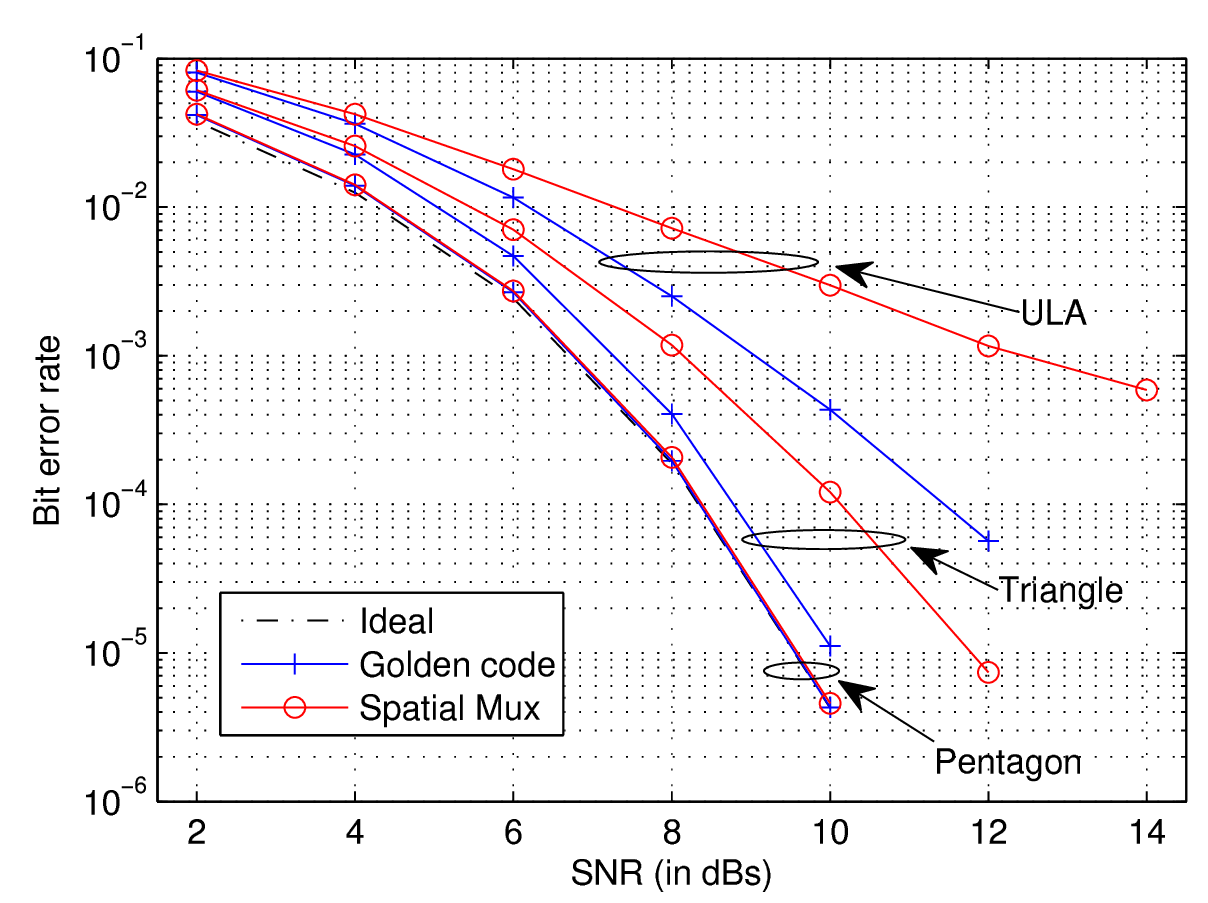}
\vspace{-3mm}
\caption{Performance of different tx arrays with tetrahedral rx array.}
\label{fig:manuscript2}
\end{figure} 

The above error performance is succinctly captured by the coding gain \mbox{$\min_{\Delta{\bf X}} {\sf d}(\mu,\Delta{\bf X})$} shown in Fig.~\ref{fig:coding_gain} as a function of $\mu$. From Example~\ref{ex:pentagon}, \mbox{$\mu \leq {2}/{3}$} for the new antenna geometry. From Fig.~\ref{fig:coding_gain} we see that the coding gains of SM and the Golden code are both equal to $1$ for all \mbox{$\mu \leq {1}/{2}$} and are larger than the SIMO coding gain for \mbox{$\mu \leq {2}/{3}$}, which explains their superiority to SIMO. On the other hand, the coding gain for linear and rectangular arrays is \mbox{$\min_{\Delta{\bf X}} {\sf d}(1,\Delta{\bf X})$}. For \mbox{$\mu=1$}, from Fig.~\ref{fig:coding_gain} we observe that SIMO has the largest coding gain followed by the Golden code and then SM. The error performances in Fig.~\ref{fig:manuscript1} show this same trend for the rectangular array at high ${\sf SNR}$.

Fig.~\ref{fig:manuscript2} compares the performance of different transmit array geometries when a tetrahedral array is used at the receiver. The \mbox{$n_t=2$} case (ULA) performs poorly since \mbox{$\mu_{\max}=1$}. While the triangular array with the Golden code achieves most of the available gain, the pentagonal array has near ideal performance.

\subsection{Error Performance with large number of receive antennas}

\begin{figure}[!t]
\centering
\includegraphics[width=3.4in]{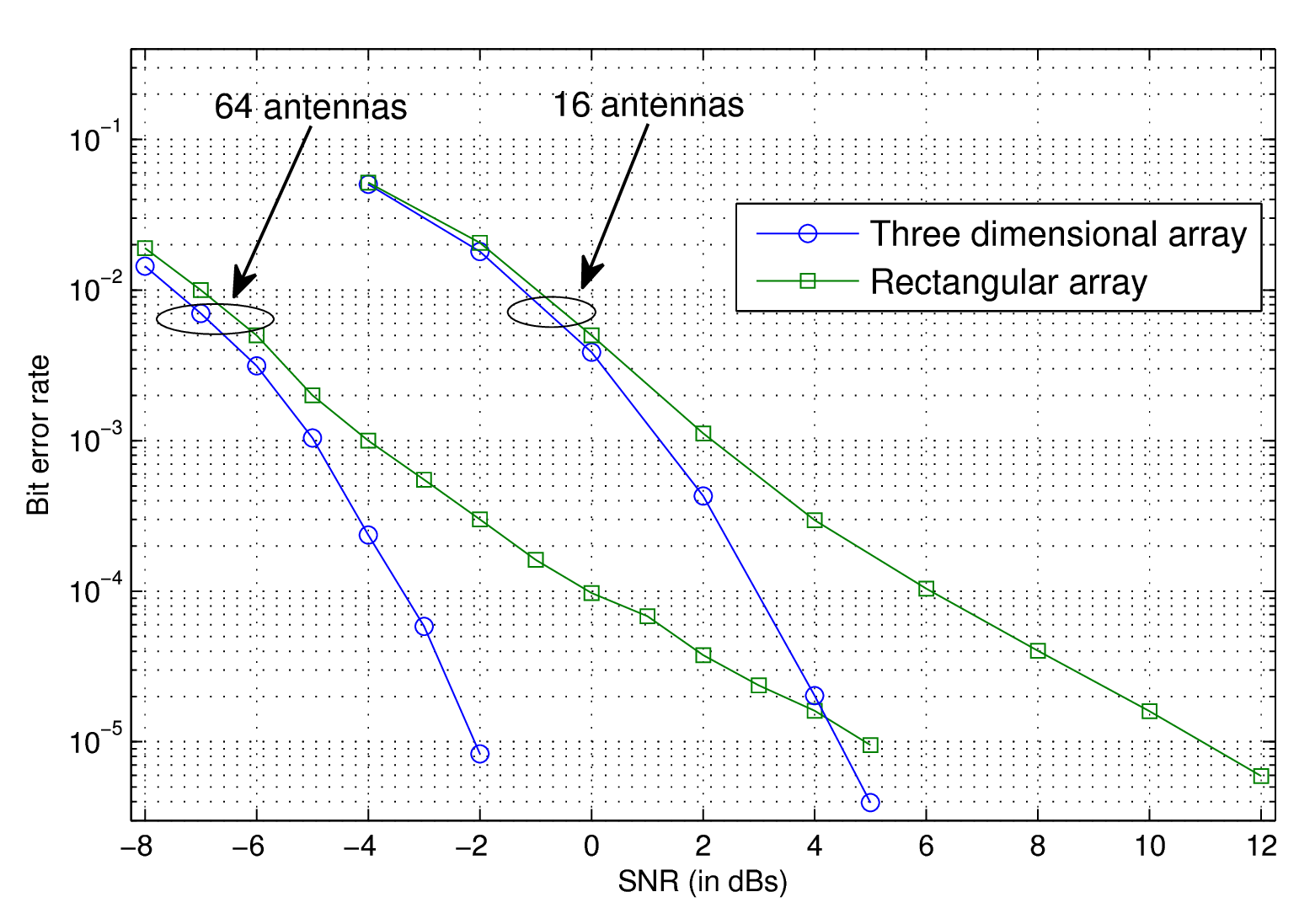}
\vspace{-3mm}
\caption{Error probability of spatial multiplexing with triangular transmit array when the receive array is {\it (i)} three-dimensional, and {\it (ii)} rectangular. Results are shown for $n_r=16$ and $n_r=64$ antennas.}
\label{fig:manuscript3}
\end{figure} 

The LoS MIMO system analysed in Section~\ref{sec:4} employs the tetrahedral receive array -- a three-dimensional antenna array for \mbox{$n_r=4$} antennas -- to enable smaller error rates than planar arrays. The geometry of the receive array is relevant even if the number of receiving antennas $n_r$ is large. Theorem~\ref{thm:planar_lower_bound} and Example~\ref{ex:vblast} show that the probability of error of the SM scheme is lower bounded up to a constant factor by ${\sf SNR}^{-3}$ for any value of $n_r$, if a planar receive array is used. On the contrary, from Example~\ref{ex:vblast:2}, the SM scheme can achieve exponential rate of decay of error probability if $n_r=4$ antennas are placed at the vertices of a regular tetrahedron. It follows that for any $n_r \geq 4$, a careful three-dimensional arrangement of $n_r$ antennas can ensure that the error rate is exponential in ${\sf SNR}$. 
For instance, if the three-dimensional arrangement includes a subset of $4$ antennas that form a tetrahedron, it immediately follows from Example~\ref{ex:vblast:2} that a sub-optimal decoder that bases its decision only on the signals received by these $4$ antennas achieves exponential error rate. Hence, the optimal ML decoder that utilizes all the $n_r$ receive antennas achieves an exponential error probability as well.

Fig.~\ref{fig:manuscript3} compares the error performance of SM scheme under planar and three-dimensional receive antenna arrays when $n_r=16,\,64$. A triangular array is used at the transmitter, $4$-QAM is chosen as the modulation scheme and ML decoding is performed at the receiver. For both values of $n_r$, we consider a URA (rectangular arrangement of receive antennas) for the planar arrangement of antennas. The three-dimensional array is chosen as a set of $n_r$ points on the surface of a sphere so that the minimum distance between the points is large. 
A table of such arrangements of points, which are known as \emph{spherical codes}, is available online~\cite{Slo_NJA}.
For fairness, the diameter of the sphere is set equal to the width of the rectangular array. The coordinates of the $n_r$ points on the sphere were obtained from~\cite{Slo_NJA}. 
As with previous simulations, we set the values of $d_t,\lambda,R_{\max}$ and $R_{\min}$ as in Example~\ref{ex:pentagon}. The inter-antenna distance $d_r$ of the URA is chosen to be $12.5$~cm when $n_r=16$ and to be $6.25$~cm when $n_r=64$. This is the optimal inter-antenna distance for the URA when the transmit and receive arrays are oriented broadside to each other and the inter-terminal distance $R=7.14$~m~\cite{BOO_Eur_07}.

It is evident from Fig.~\ref{fig:manuscript3} that array geometry is an important design parameter even when $n_r$ is large. 
The error rates of rectangular arrays shown in Fig.~\ref{fig:manuscript3} decay as ${\sf SNR}^{-2}$ at high ${\sf SNR}$.
The gain due to the three-dimensional array is about $7$~dB at an error rate of $10^{-5}$ for both $n_r=16$ and $64$.

\section{Conclusion} \label{sec:conclusion}

We studied the error performance of arbitrary coding schemes in $2 \times n_r$ LoS MIMO channels where the communicating terminals have random orientations. We analyzed the effects of some receive array geometries on error probability, and showed that, unlike linear, circular and rectangular arrays, the error rate with a tetrahedral array decays faster than that of a rank $1$ channel. Using tetrahedral and polygonal arrays, we designed a LoS MIMO system that provides a good error performance for all transmit and receive orientations.
By modelling the ${\bf R}$ matrix, we derived error probability bounds for the case when the number of transmit antennas used for signalling is $2$. Analysis of the performance when more than $2$ transmit antennas are used is yet to be addressed.

%%%% Acknowledgements %%
%% \section*{Acknowledgment}
%% The authors would like to thank the anonymous reviewers and the editor for their constructive comments that have improved the content and the presentation of this paper.

%%%% references %%%%%
%% \bibliography{IEEEabrv,VT-2016-00161_Final}


\end{document}